\documentclass[lettersize,journal]{IEEEtran}

\usepackage{tikz}
\usepackage{textcomp}
\usepackage{hyperref}
\usepackage{lipsum}

\usepackage{amsmath,amsfonts,amssymb}
\usepackage{algorithmic}
\usepackage{algorithm}
\usepackage{array}
\usepackage[caption=false,font=normalsize,labelfont=rm,textfont=rm]{subfig}
\usepackage{textcomp}
\usepackage{stfloats}
\usepackage{url}
\usepackage{verbatim}
\usepackage{graphicx}
\usepackage{cite}
\usepackage{xcolor}
\usepackage{tabularx}
\usepackage{booktabs}

\usepackage{subfig}
\usepackage{hyperref}
\hypersetup{
    hidelinks,
    }

\newcommand\copyrighttext{%
  \large\sf\textcopyright~2023 IEEE. Personal use of this material is permitted.
  Permission from IEEE must be obtained for all other uses, in any current or future
  media, including reprinting/republishing this material for advertising or promotional
  purposes, creating new collective works, for resale or redistribution to servers or
  lists, or reuse of any copyrighted component of this work in other works. 
  DOI: \href{https://doi.org/10.1109/TITS.2023.3285296}{10.1109/TITS.2023.3285296}}
\newcommand\copyrightnotice{%
\begin{tikzpicture}[remember picture,overlay]
\node[anchor=south,yshift=100pt] at (current page.south) {\fbox{\parbox{\dimexpr\textwidth-\fboxsep-\fboxrule\relax}{\copyrighttext}}};
\end{tikzpicture}%
}

\begin{document}

\onecolumn

\copyrightnotice

\vfil

{\Large \noindent\sf Th.~Apostolakis, M.A.~Makridis, A.~Kouvelas, K.~Ampountolas, ``Energy-based Assessment and Driving Behavior of ACC Systems and Humans Inside Platoons," \emph{IEEE Transactions on Intelligent Transportation Systems}, DOI: \href{https://doi.org/10.1109/TITS.2023.3285296}{10.1109/TITS.2023.3285296}.}

\bigskip \bigskip \bigskip \bigskip \bigskip 

{\Large \noindent\sf The material cannot be used for any other purpose without further permission of the publisher and is for private use only.}
\bigskip

\bigskip \bigskip \bigskip 
{\Large \noindent \sf
There may be differences between this version and the published version. You are advised to consult the publisher’s version if you wish to cite from it.}

\twocolumn
\clearpage

\title{Energy-based Assessment and Driving Behavior of ACC Systems and Humans Inside Platoons}

\author{Theocharis Apostolakis, Michail A. Makridis, Anastasios Kouvelas, and Konstantinos Ampountolas,~\IEEEmembership{Member,~IEEE}
\thanks{This paper has supplementary downloadable material available at \url{https://github.com/TheocharisAp/ACC_Energy_Assessment}, provided by the authors.} 
\thanks{This work was supported by the Center of Research, Innovation and Excellence  (CRIE) of the University of Thessaly, under the project ``campaigningACC".}
\thanks{Th.~Apostolakis and K.~Ampountolas are with the Automatic Control and Autonomous Systems Laboratory, Department of Mechanical Engineering, University of Thessaly,  38334 Volos, Greece (e-mail: \{tapostolakis, k.ampountolas\}@uth.gr).}
\thanks{M.~A.~  Makridis and A.~Kouvelas are with the Institut f{\" u}r Verkehrsplanung und Transportsysteme (IVT), ETH Z{\" u}rich, 8093 Z{\" u}rich, Switzerland (e-mail: \{mmakridis, kouvelas\}@ethz.ch).}
}

\markboth{IEEE Transactions on Intelligent Transportation Systems}%
{Shell \MakeLowercase{Apostolakis \textit{et al.}}: Energy-based Assessment and Driving Behavior of ACC Systems and Humans Inside Platoons}

\IEEEpubid{0000--0000/00\$00.00~\copyright~2023 IEEE}

\maketitle

\begin{abstract}
Evidence in the literature shows that automated and human driving modes demonstrate different driving characteristics, i.e., headway policy, spacing policy, reaction time, comfortable acceleration, and others. These differences alter observed traffic dynamics and have an impact on energy consumption. This paper assesses the energy footprint of commercially implemented adaptive cruise control (ACC) systems and human drivers in car-following formation via different models using empirical observations on very similar driving cycles and/or routes. Most importantly, it initiates a critical discussion of the findings under the behavioral properties of each mode. Findings show that: ACC systems propagate an increasing energy consumption upstream, while human drivers do not; they succeed in maintaining a constant time-headway policy, operating very reliably; they develop a strong bond with their leader compared to their human counterparts; the two modes (humans and ACCs) are operating in different phase-space areas with room for improvement. Overall, findings show that ACC systems must be optimized to achieve a trade-off between functional requirements and eco-driving instructions.
\end{abstract}


\section{Introduction}\label{sec:Introduction}

\IEEEPARstart{R}{oad} transportation is responsible for a large share of greenhouse gas emissions (GHG) worldwide \cite{EPA}. At the same time, automated driving will change how people move in the existing networks, introducing new driving behaviors. Vehicle automation promises to increase road safety, network capacity, and traffic flow, currently limited due to heterogeneity in vehicle dynamics and human driving behaviors \cite{10.1186/s12544-020-00407-9}. This creates a debate in the literature regarding their energy footprint compared to human drivers \cite{WADUD20161}. Only a few studies provide insights into the energy consumption differences between partially automated vehicles and human drivers based on detailed trajectories. A fair comparison requires a similar driving cycle for both modes, and such data observations are scarce.

Let us first briefly discuss the operational principles of Adaptive cruise control (ACC) systems, responsible for their possible different energy footprint than human drivers. We have limited information on the precise operation of deployed ACC systems in commercial vehicles due to the proprietary rights of the manufacturers. However, the basic operating principles are inferred through observations of such systems on the road. Vehicles equipped with ACC have several sensors onboard, such as cameras, radars, or LiDAR, to monitor the surrounding environment and regulate the vehicle's speed accordingly. The ACC detects the preceding vehicle's position and speed regularly. Then it adjusts the speed of the ego-vehicle based on several factors (some unknown) such as the spacing from the preceding vehicle, the speed difference between the preceding and following vehicles, the desired speed, and a desired target time-headway. The system disengages at very low speeds for some vehicles or in the case of an imminent collision when other systems like Automated Emergency Braking might take over. 

Although ACC systems, a proxy for automated driving, were initially expected to increase the efficiency of traffic networks, it soon became apparent that this is highly unlikely. As the penetration rate of ACC-equipped vehicles increases, literature studies analyze empirical data and report escalating concerns on undesired properties, such as string instability, safety concerns, negative impact on road capacity, high energy demand and fuel consumption \cite{Ioannou_Chen, Swaroop, Swaroop_stability, euroFOT:2012, liang_stability, Shi2022}. Most of the works in the literature concentrate on either positive (earlier works) or negative (recent works) argumentation around ACC systems. Herein, we provide an energy-based analysis along with a critical discussion on the driving behavior of such systems for an overview of the topic, highlighting both observed downsides but also benefits related to commercial ACC systems. The analysis is designed on similar driving cycles from experimental campaigns with platoons driven by humans and ACC systems, thus enabling a fair comparative assessment \cite{MAKRIDIS2021103047, article}.

\IEEEpubidadjcol

Studies on the energy footprint of ACC-equipped vehicles are based on traffic simulation \cite{MAKRIDIS2020117399} or empirical data \cite{SHI2022103253} obtained from experimental campaigns. An experimental study with a fleet of over 50 vehicles driven for almost 200.000 miles concludes positively about the induced fuel consumption of commercial ACC systems, as they could reduce GHG emissions at low speeds \cite{GeneralMotorsLLC}. On the contrary, another experimental campaign with platoons of ACC-equipped vehicles in a test site concludes that energy-wise these systems are inefficient. Tested ACCs were found string unstable; therefore, small perturbations were amplified upstream of the platoon, leading to increasing (upstream) fuel consumption for the same route \cite{CIUFFO2021103305}. Both of the above studies have valid conclusions but counterarguments can be easily found. ACC systems operate most commonly at high speeds; so, any benefit at low speeds might never be realized. Furthermore, forming ACC platoons is not very common, especially in mixed traffic conditions (humans and automated drivers). The higher energy consumption derives from the amplification of perturbations, which are not so common at high speeds where ACCs usually operate. 

Another field study focuses on the maximum number of ACC-equipped vehicles that can form a platoon (theoretically infinite). It concludes that platoon formations of more than typically three to four vehicles are problematic because of instabilities in the car-following behavior. Hence, discomfort and large fuel consumption \cite{doi:10.1177/0361198119845885}. Finally, a study on the Vicolungo campaign of the OpenACC dataset \cite{article} shows that from individual and platoon perspectives, ACC followers tend to have higher energy consumption than their human counterparts, questioning the positive impact of ACC systems on fuel and energy consumption.

In the present work, the energy footprints of ACC systems and human drivers are assessed by four state-of-the-art energy demand \cite{Book:Vehicles} and fuel consumption models, namely VT-micro \cite{ahn1998microscopic}, VSP \cite{JimnezPalacios1999UnderstandingAQ, DUARTE2015251}, and ARRB \cite{ARRB1986}. The parameters of the energy estimation models have been derived in different periods and calibrated on fleets with different power dynamics. Therefore, the absolute quantitative results yield different values. Nevertheless, the energy tendency observed reveals interesting findings summarized in this work. The assessment is performed on empirical data from two independent experimental car-following campaigns on similar driving cycles with and without using ACC \cite{MAKRIDIS2021103047,article}. The study expands to a broader critical discussion around the energy footprint of ACC systems, trying to create a link with their driving characteristics compared to human drivers. As will be shown later, the reasons vary; they derive from ACC operation and behavior, and even the priorities (functional specifications) that the manufacturers set during ACC systems' design.

The present paper shows that: (a) ACC systems are less energy efficient than human drivers inside platoons; (b) ACC driving operation (strong accelerations, steep speeds) and its impact (speed overshoots, string instabilities) may negatively affect ACC systems' energy footprint; (c) ACC functional specifications (i.e., constant time-headway policy, distribution of time and space-gap values, bond between the ACC participants) are set in the first place for ACC manufacturers, neglecting ACC systems' energy efficiency; (d) Humans and ACC driving modes are operating in different phase-space areas, thus there is room of improvement for both agents; (e) ACC systems must be optimized to achieve a trade-off between functional requirements and eco-driving instructions.  

The paper is structured as follows. Section~\ref{sec:Methods} presents four well-established energy, and fuel consumption models. Section~\ref{sec:App_Res} presents two heterogeneous experimental car-following testing campaigns employing different data acquisition methods, design, and execution. It also assesses the energy footprint of ACC and human drivers as obtained from the application to the considered energy and fuel consumption models. Section~\ref{sec:discussion} critically discusses and analyzes the energy footprint of stock ACC systems and their merits concerning functional specifications. Section~\ref{sec:Conclusion} summarises the main findings of this work and offers suggestions for future work.

\section{Methodology}\label{sec:Methods}

Four independent energy demand and fuel consumption models are used here to analyze the energy behavior of the ACC systems. The first model computes the tractive energy demand on the wheels \cite{Book:Vehicles, article}, ruling out the effect of the propulsion system. The other three models, namely VT-micro \cite{ahn1998microscopic}, VSP \cite{JimnezPalacios1999UnderstandingAQ, DUARTE2015251}, and ARRB \cite{ARRB1986}, estimate the instantaneous fuel consumption. 

\subsection{Tractive Energy Consumption}

Tractive energy consumption considers only the tractive power demand on the vehicle's wheels without considering the powertrain dynamics and the regenerative braking power. The outputs of this model do not represent energy consumption. Still, a vehicle-agnostic energy indication creates a fair comparison between driving modes, i.e., human and automated driving \cite{article}. The instantaneous tractive power, $P_t$ [kW] at time $t$, required to move the engine at the defined velocity and surpass the aerodynamic and rolling resistances, is given by \cite{Book:Vehicles}:\footnote{Notation: $[\,\cdot\,]^+ \triangleq \max\{0,\cdot\}$}  
\begin{equation}\label{eq:Tractive_Energy}
   P_t = 
10^{-3}[(F_0 + F_1v_t + F_2v_t^2+ 1.03ma_t + mg \sin\theta)v_t ]^+,
\end{equation}
where $F_0= 213$ N, $F_1= 0.0861$ N$\cdot$s/m, and $F_2= 0.0027$ N$\cdot$s$^2$/m$^2$, are road load coefficients that describe the relationship between overall resistances to motion and the vehicle's speed; $m$ [kg] is the vehicle's mass; $v_t$ [m/s] and $a_t$ [m/s$^2$] are the ego vehicle’s speed and acceleration at time $t$, respectively; $\theta$ [rad] is the road gradient; and $g$ is the gravitational acceleration. 

Integration over time of the instantaneous tractive power, $P_t$ [kW], and division by the total distance traveled, results to the vehicle’s tractive energy consumption, $E_c$ [kWh/100 km] \cite{article}:\begin{equation}\label{eq:Tractive_Energy_new}
    E_c = \displaystyle \int_0^T P_t \, dt \Bigg/ \Bigg(0.036  \displaystyle\int_0^T v_t\, dt\Bigg),
\end{equation}
where $dt$ is the time interval between consecutive measurement points, and $T$ [s] denotes the total travel time. The factor $0.036$ in the denominator of \eqref{eq:Tractive_Energy_new} is applied so that the results are available in kWh/100 km.

\subsection{Fuel Consumption Models}

The impact of ACC driving behavior on fuel consumption is assessed by three fuel consumption models, namely, VT-micro \cite{ahn1998microscopic}, VSP \cite{DUARTE2015251}, and ARRB \cite{ARRB1986}, focusing on the instantaneous fuel consumption.

\subsubsection{The VT-micro model}

The Virginia Tech (VT)-micro model is a microscopic dynamic emission, and fuel consumption model \cite{ahn1998microscopic}. This model emerged after experimentation with different polynomial approximators of speed and acceleration profiles \cite{articleRa}. The instantaneous fuel consumption, $F_t$[L/s] at time $t$, of a vehicle can be expressed as \cite{ahn1998microscopic}:
\begin{equation}\label{eq:VT_micro}
F_t= \exp\Bigg(\sum_{i=0}^{3} \sum_{j=0}^{3} K_{{i}{j}} (v_t)^{i} (a_t)^{j}\Bigg),
\end{equation}
where $v_t$ [m/s] and $a_t$ [m/s$^2$] are the speed and acceleration of the vehicle at time $t$, respectively, and  $K_{{i}{j}}$ are constant coefficients that can be found in Table \ref{tab:Coef} \cite{ZEGEYE2013158}.

\newcommand{\minus}{\scalebox{0.75}[1.0]{$-$}}
\begin{table}[tbp]
    \caption{The VT-micro Model Coefficients.}\label{tab:Coef}
    \centering
        \begin{tabular}{c|cccc}
            \toprule
            $K_{{i}{j}}$ & $j =$ 0 & $j =$ 1 & $j =$ 2 & $j =$ 3 \\
            \midrule
            $i =$ 0 & \minus7.537 & 0.4438 & 0.1716 & \minus0.0420 \\
            $i =$ 1 & 0.0973 & 0.0518 & 0.0029 & \minus0.0071 \\
            $i =$ 2 & \minus0.003 & \minus7.42E\minus04 & 1.09E\minus04 & 1.16E\minus04 \\
            $i =$ 3 & 5.3E\minus05 & 6E\minus06 & \minus1E\minus05 & \minus6E\minus06 \\
            \bottomrule
        \end{tabular}
\end{table}

\subsubsection{The VSP model}

The vehicle specific power (VSP) model is defined as the instantaneous engine power demand required to overcome the rolling resistance and aerodynamic drag and to increase the vehicle's kinetic and potential energy, divided by the mass of the vehicle \cite{JimnezPalacios1999UnderstandingAQ}. For this study, the following form is considered:  
\begin{equation}\label{eq:5}
    P_t = v_t(1.1a_t + 9.81\theta + 0.132) + 3.02\times 10^{-4}  v_t^3,
\end{equation}
where $P_t$ [W/kg] is the instantaneous VSP at time $t$; $v_t$ [m/s] and $a_t$ [m/s$^2$] are the speed and acceleration of the vehicle at time $t$, respectively; $\theta$ [rad] is the road grade. 

VSP is calibrated with a portable laboratory collecting on-road measurements, i.e., fuel consumption, pollutant emissions, and vehicle dynamics of nineteen EURO 5 vehicles \cite{DUARTE2015251}. For each time instant, according to the power demand resulting from the vehicle specific power, $P_t$ [W/kg], the corresponding VSP mode was calculated; VSP values are divided into mode bins consisting of 1 W/kg. The fuel consumption trend is approximated as a function of VSP modes through six coefficients adjustable per vehicle according to the certification inputs \cite{DUARTE2015251}. The instantaneous fuel consumption, $F_t$ [g/s] at time $t$, can be expressed as a function of vehicle specific power, $P_t$ [g/s], as follows: 
\begin{align}\label{eq:6}
     F_t(P_t) = \begin{cases} f, & \text{if \;$P_t < -10$},\\ \alpha P_t^2 + \beta P_t + \gamma, & \text{if \;$-10 \leq P_t \leq 10$},\\ \delta P_t + \varepsilon, & \text{if \; $P_t \geq 10$},\end{cases}
\end{align}
The parameter values that we use uniformly for all the vehicles are $f = 2.48{\rm E}\minus03$, $\alpha= 1.98{\rm E}\minus03$, $\beta =3.97{\rm E}\minus02$, $\gamma = 2.01{\rm E}\minus01$, $\delta = 7.93{\rm E}\minus02$, and $\varepsilon = 2.48{\rm E}\minus03$; coefficients corresponding to one of the testing vehicles \cite{DUARTE2015251}. To obtain the results in L/s, the instantaneous fuel consumption, $F_t$ [g/s], is divided by the fuel density. In this analysis, assuming only diesel vehicles, the oil density $\rho$ is estimated at around 850 g/L.

\subsubsection{The ARRB model}

The Australian road research board (ARRB) model is an instantaneous fuel consumption model well suited for determining the incremental effects on fuel consumption resulting from changes in traffic flow  \cite{ARRB1986}. The fuel consumption, $F_t$ [ml/s] at time $t$, is expressed as the following polynomial function of instantaneous speed and acceleration:
\begin{equation}\label{eq:7}
    F_t = \beta_1 + \beta_{2} v_t + \beta_{3} v_t^2 + \beta_{4} v_t^3 + \gamma_{1} v_t a_t + \gamma_{2} v_t ([a_t]^+)^2, 
\end{equation}
where $v_t$  [m/s] and $a_t$ [m/s$^2$] are the speed and acceleration of the vehicle at time $t$, respectively; The parameters values adopted for all vehicles in this study are $\beta_1 = 0.666$; $\beta_2 = 0.019$; $\beta_3 = 0.001$; $\beta_4 = 0.00005$; $\gamma_1 = 0.12$; $\gamma_2 = 0.058$.

\subsubsection{Fuel consumption estimation}

To obtain meaningful results from the fuel consumption models, the vehicle's fuel consumption of the various models, $F_c$ [L/100 km], results by integrating the instantaneous fuel consumption, $F_t$ [L/s], and dividing it by the distance covered, which corresponds to the integration of the instantaneous speed over time:
\begin{equation}\label{eq:8}
    F_c = \int_0^T F_t \, dt \Bigg/ \Bigg(10^{-5}\displaystyle\int_0^T v_t\, dt\Bigg),
\end{equation}
where $dt$ is the time interval between consecutive measurement points, and $T$ [s] denotes the total travel time. The factor of 10$^{-5}$ in the denominator is applied so that the results are available in L/100 km.

\subsection{Platoon Normalization}\label{sec:Normalization}

Results presented here assess the energy impact of ACC systems from a platoon perspective. Similar traffic conditions and driving cycles for all the vehicles in the platoon ensure a fair comparison but the vehicles are of different brands or models, and thus, they have different specifications. A scaling technique presents results focused on driving behaviors independently of other factors. Specifically, we assume the same parameter values for the vehicle's specifications, road load coefficients, and mass among the various energy and fuel consumption models. This process leads to consumption values that do not reflect in absolute terms each vehicle but enable relative comparisons, which is the goal of this work, i.e., comparing different driving modes.

\section{Results}\label{sec:App_Res}

Data from two campaigns of the OpenACC dataset are used to compare the energy footprint of ACC-driven and human-driven vehicles \cite{MAKRIDIS2021103047}.  The Vicolungo dataset includes real-world platoon data on the same route in both directions, i.e., from Ispra to Vicolungo and back. The experiment emulates normal driving conditions (for the leader). In one direction (southbound), all five vehicles are driven by human drivers, while on the way back (northbound), the three vehicles in the middle of the platoon are driven by their ACC system. The order of the vehicles in the platoon is the same in both directions and there are no trip repetitions. This design enables a fair comparison between modes since it refers to the same route. For more information on the campaign, we refer the reader to the corresponding papers \cite{MAKRIDIS2021103047, article}. 

The AstaZero is a test track (protective environment) in Sweden, and this dataset includes observations with platoons of five vehicles \cite{MAKRIDIS2021103047}. The observations were acquired with a differential GPS device of very high accuracy. The data used here refer to the same driving cycle, one referring to human-driven vehicles, except the leader, and another referring to the same vehicles with ACC enabled. For more information on the campaign, see  \cite{MAKRIDIS2021103047}.

\begin{figure*}[tbp]
     \centering
     \subfloat[{Tractive energy consumption}\label{fig:Tr_energy_Italy}]{%
         \includegraphics[width=0.25\linewidth]{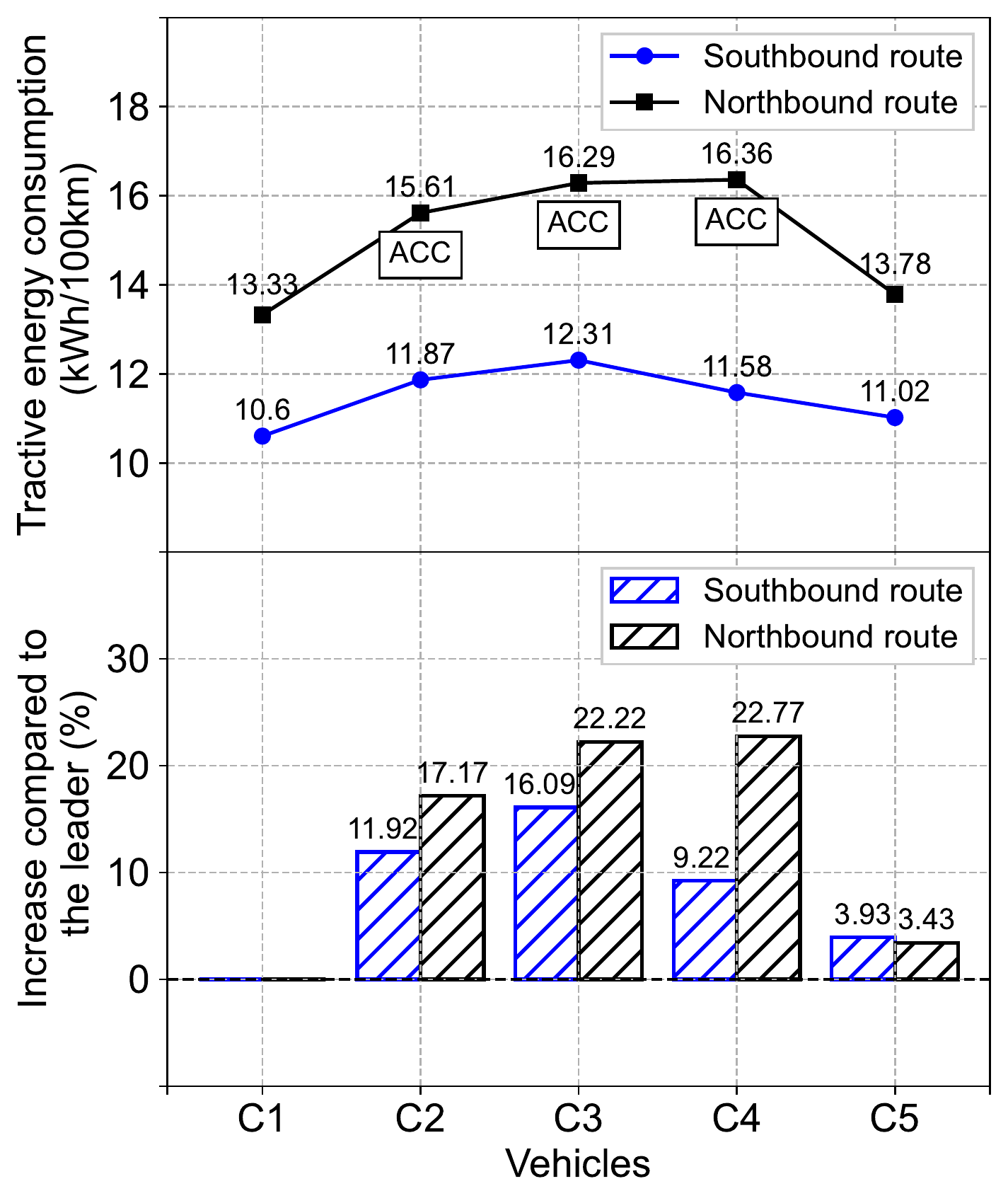}}
     \hfill
     \subfloat[Fuel consumption: VT-micro\label{fig:VT-micro_Italy}]{%
         \includegraphics[width=0.25\linewidth]{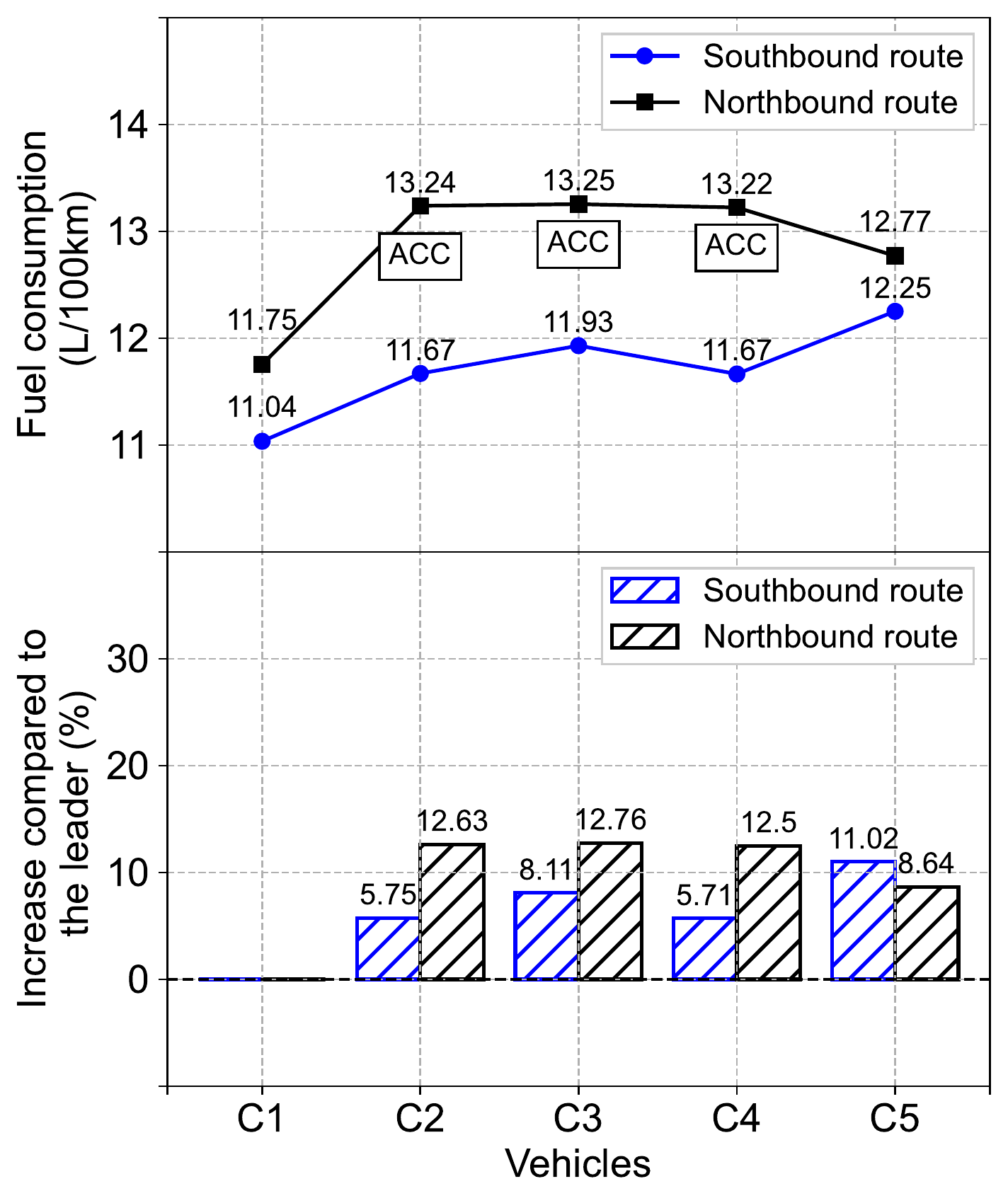}}
     \hfill    
     \subfloat[Fuel consumption: VSP\label{fig:VSP_Italy}]{%
         \includegraphics[width=0.25\linewidth]{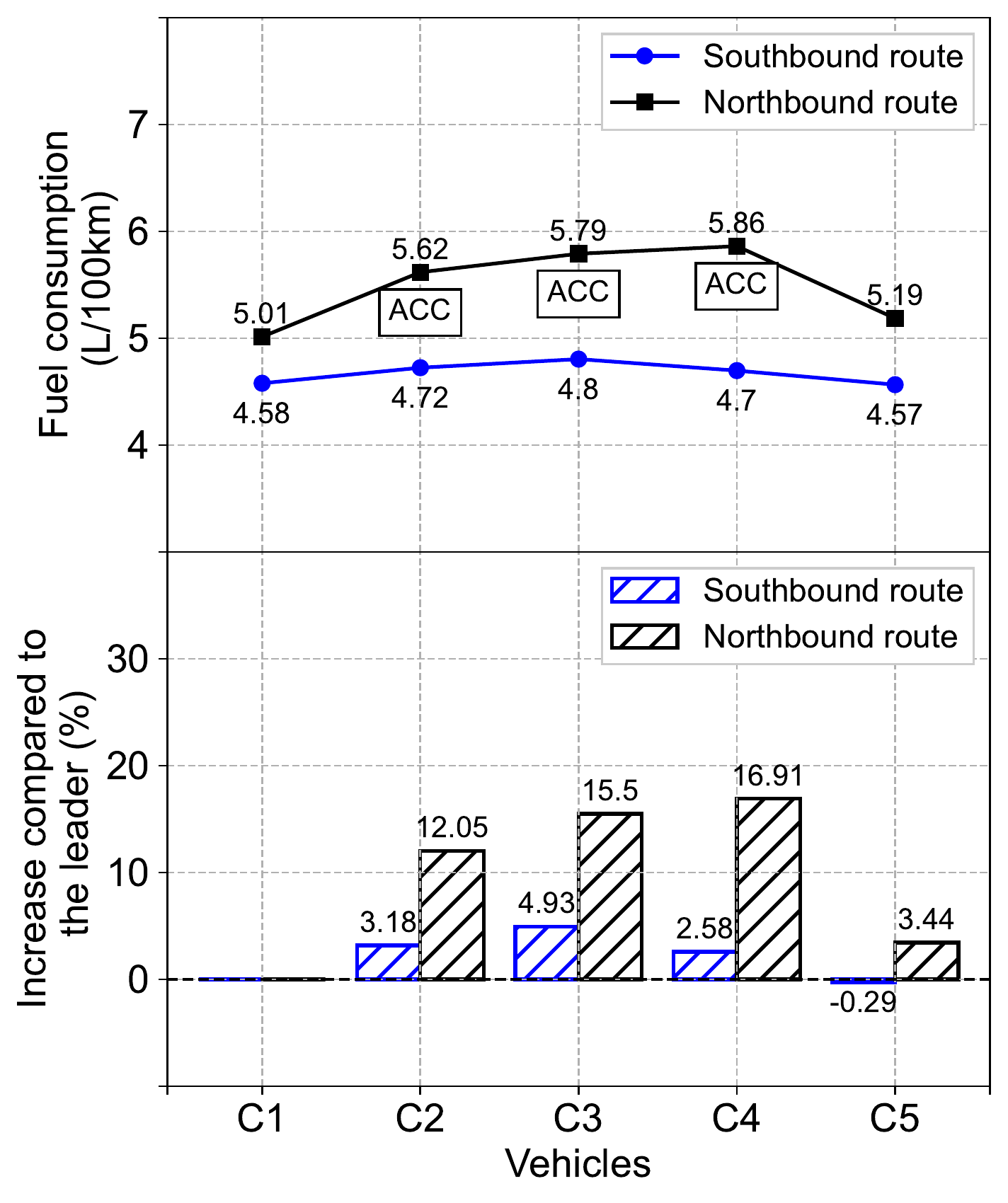}}
     \hfill
     \subfloat[Fuel consumption: ARRB\label{fig:ARRB_Italy}]{%
         \includegraphics[width=0.25\linewidth]{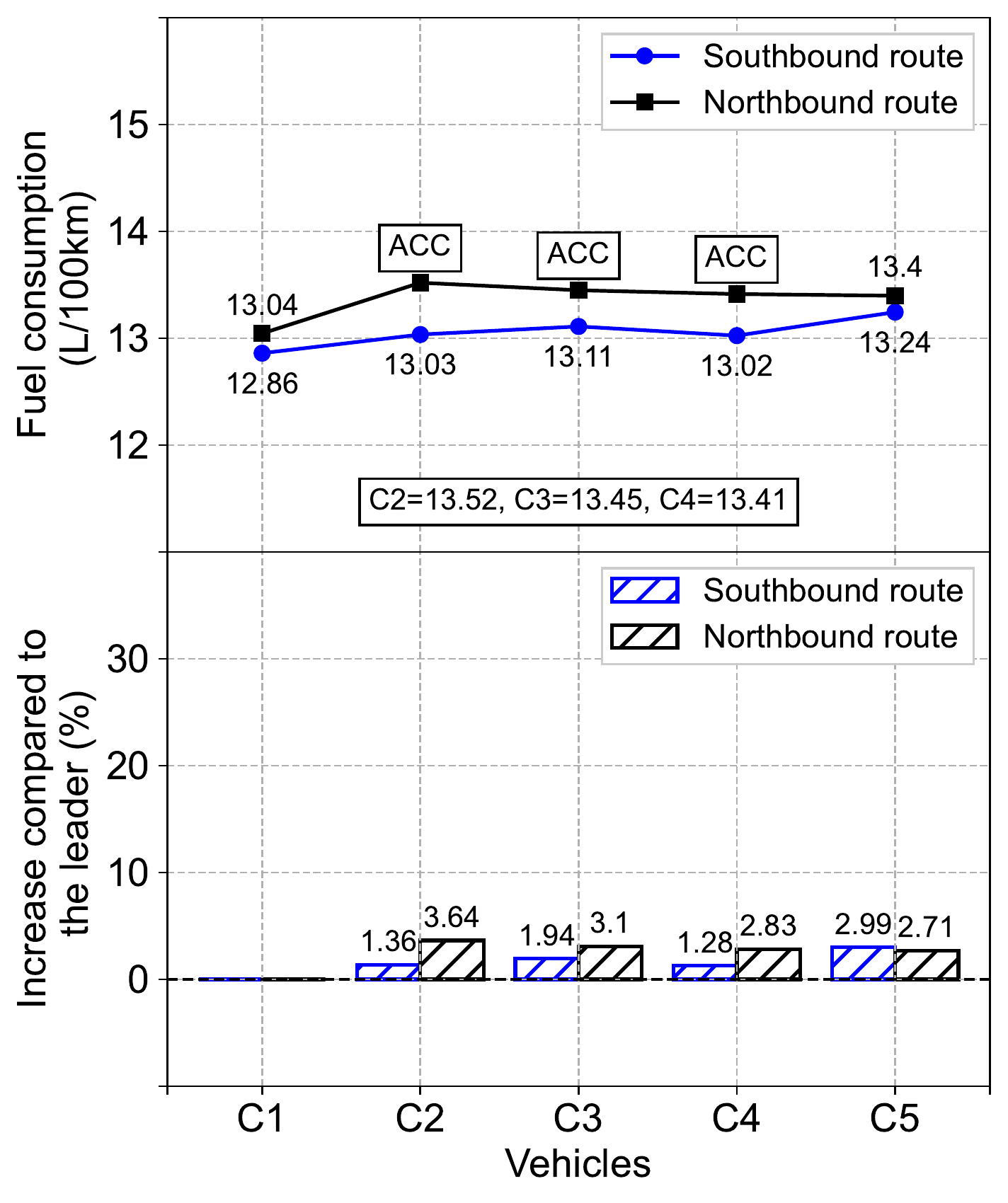}}
    \caption{Energy demand and fuel consumption for the Vicolungo campaign.}
    \label{fig:Energy/Fuel_consumption_Italy}
\end{figure*}

Fig.~\ref{fig:Energy/Fuel_consumption_Italy} presents the absolute and relative energy estimation results per model for the Vicolungo campaign. Fig.~\ref{fig:Tr_energy_Italy} reveals a tendency for increasing energy propagation upstream in the platoon with ACC-engaged in the Northbound (NB) route. The tractive energy values of the ACC followers (C2, C3, and C4) tend to increase consecutively, in contrast to those their human counterparts achieve in the Southbound (SB) route. Consequently, ACC systems appear to be less energy efficient than human drivers inside the platoon.

Figs~\ref{fig:VT-micro_Italy}--\ref{fig:ARRB_Italy} present the obtained results of fuel consumption (in L/100 km) for the three considered models. Almost the same tendency is revealed among the various fuel consumption models regarding ACC driving behavior but the amplification factor estimated by each model differs. The VT-micro model reveals a strong fuel consumption amplification by the first ACC participant (C2) in the NB route, with the rest of them (C3, C4) maintaining the values on high levels in relation to those their human counterparts achieve in the SB route. Similar results are shown by the VSP model. The last human counterpart (C5) in the NB route reveals a relatively downward value trend, in both VT-micro and VSP models, trying to absorb the increasing fuel consumption propagation occurring upstream in the platoon from the ACC participants. The same behavior is also observed in the tractive energy consumption mentioned above. Both driving modes seem to achieve very small value fluctuations in the case of ARRB model, showing steady-state fuel consumption profiles; ARRB model seems to be a more conservative model. The value ranges among the three models are significantly different, which is normal as the models are not calibrated on specific vehicles. All models agree that ACC systems in this test tend to increase fuel consumption inside the platoon and hence, are less efficient compared to human drivers. 

\begin{figure*}[tbp]
     \centering
     \subfloat[{Tractive energy consumption}\label{fig:Tr_energy_AstaZero}]{%
         \includegraphics[width=0.25\linewidth]{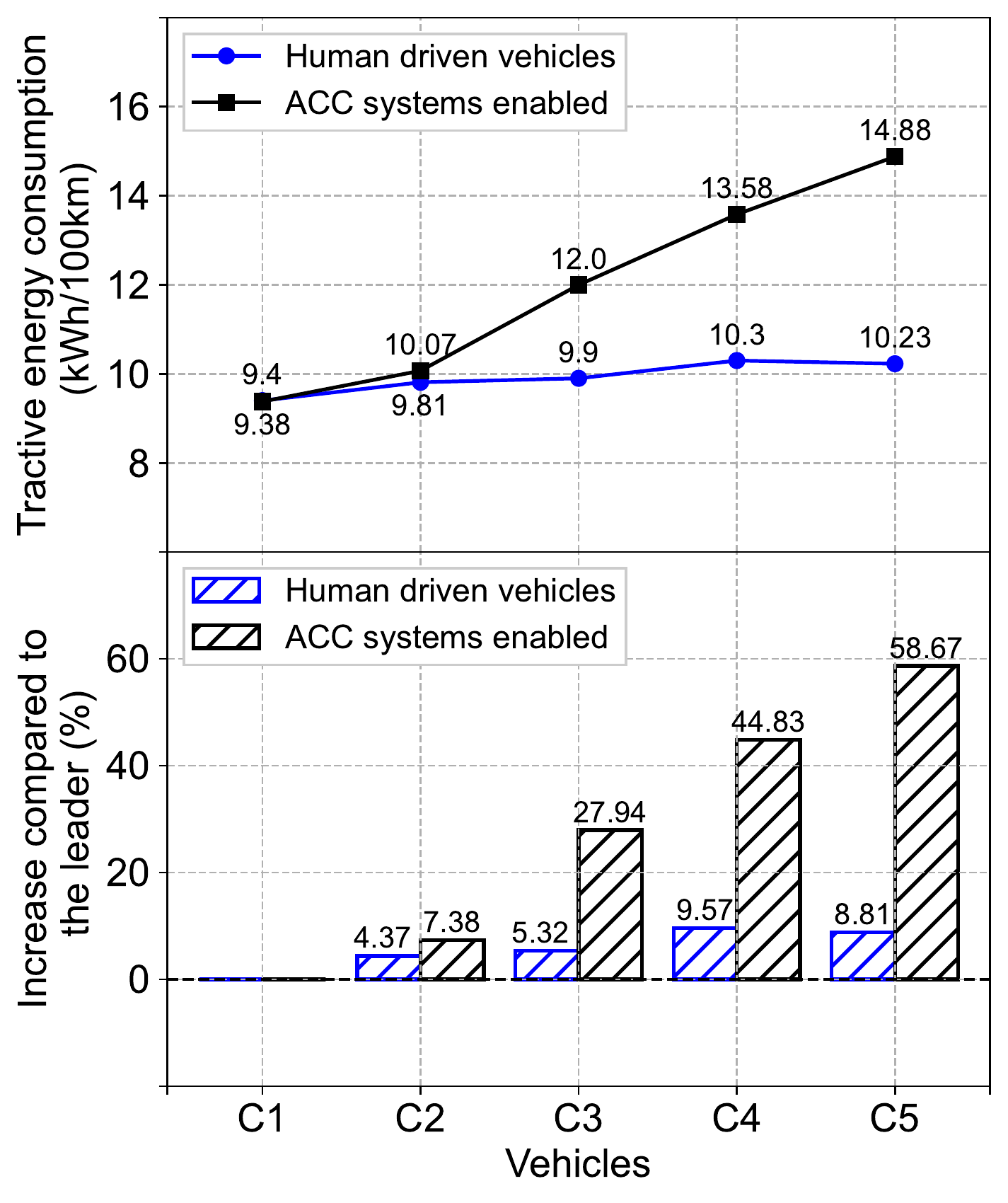}}
     \hfill
     \subfloat[Fuel consumption: VT-micro\label{fig:VT-micro_AstaZero}]{%
         \includegraphics[width=0.25\linewidth]{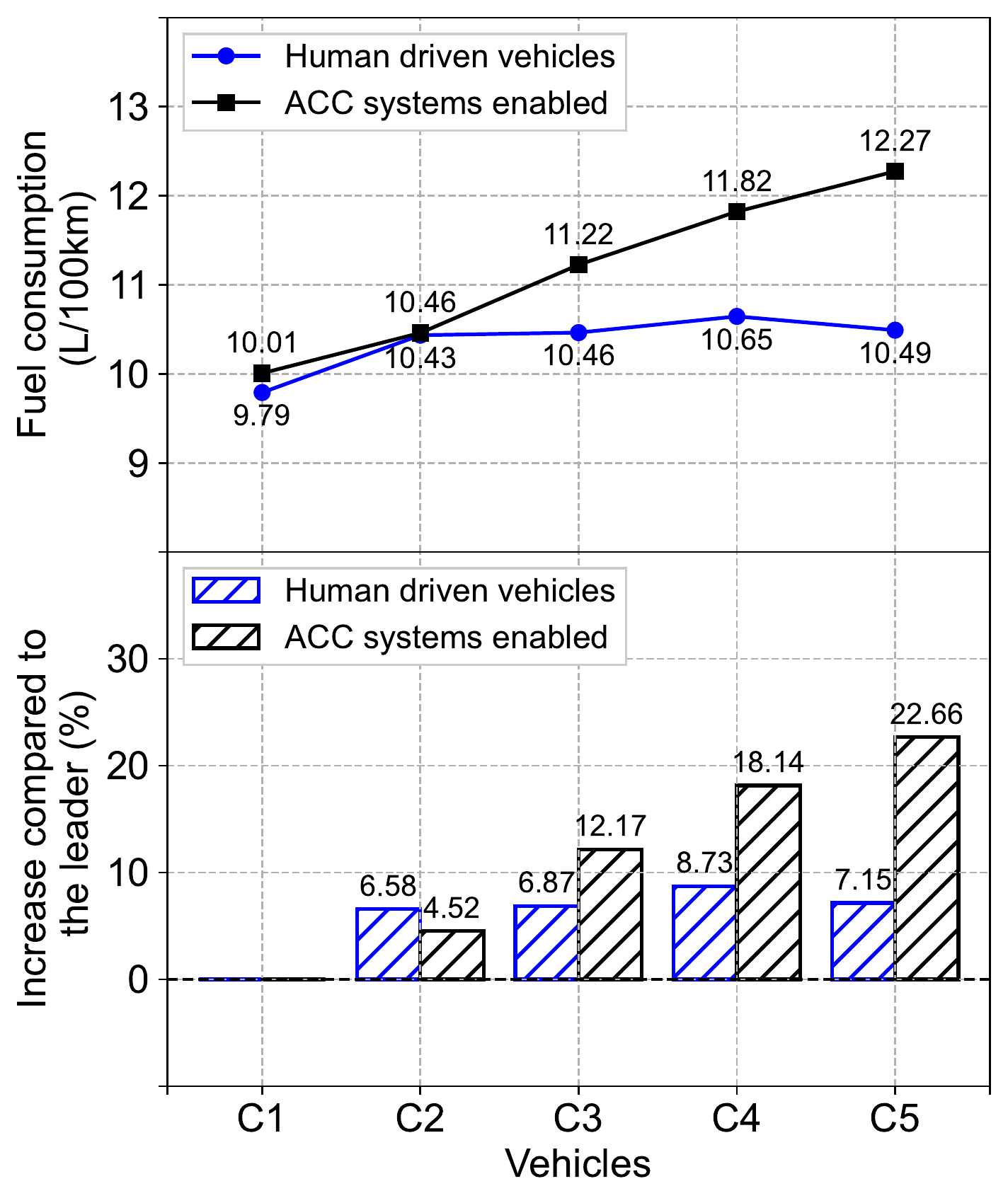}}
\hfill
     \subfloat[Fuel consumption: VSP\label{fig:VSP_AstaZero}]{%
         \includegraphics[width=0.25\linewidth]{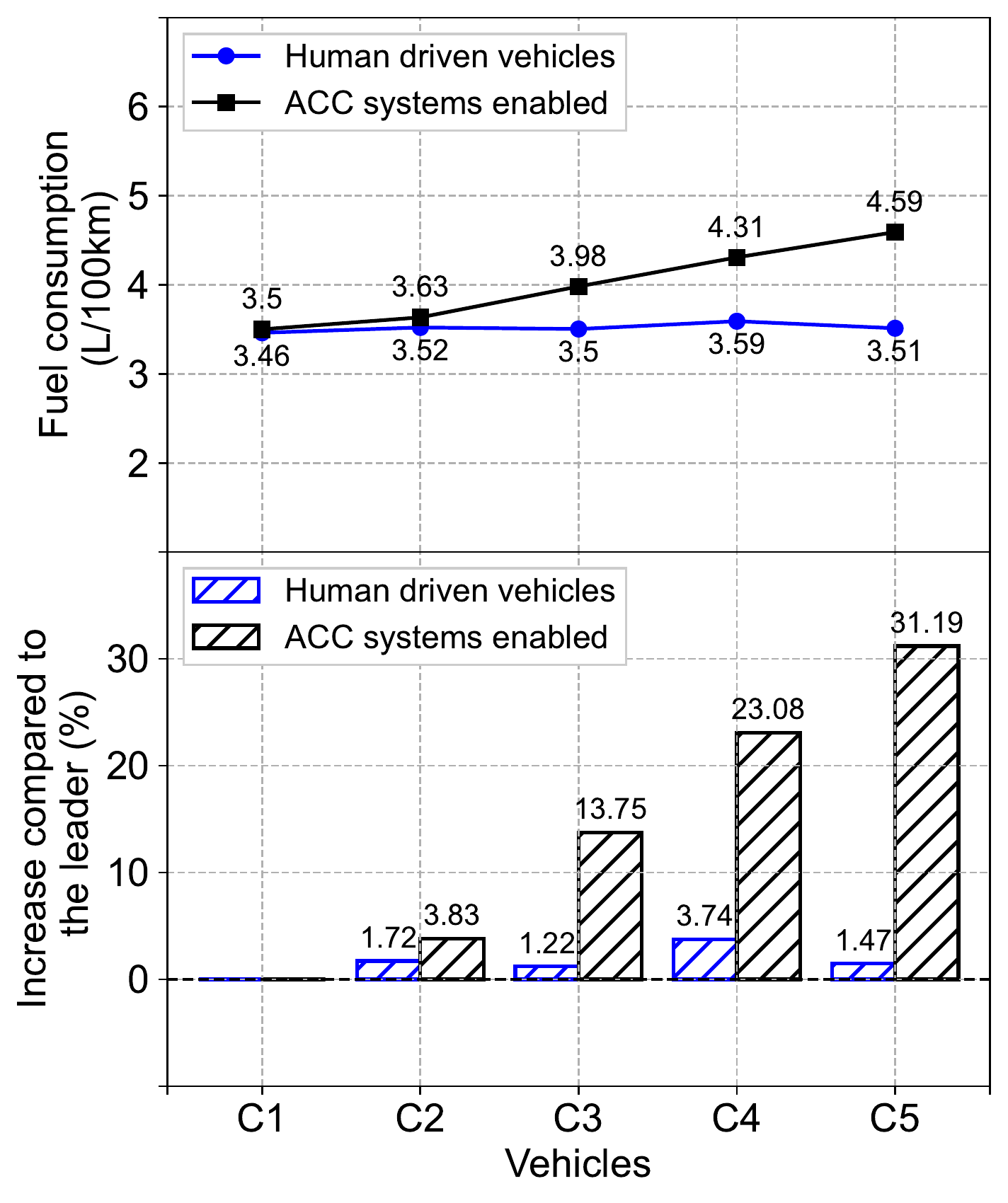}}
     \hfill
     \subfloat[Fuel consumption: ARRB\label{fig:ARRB_AstaZero}]{%
         \includegraphics[width=0.25\linewidth]{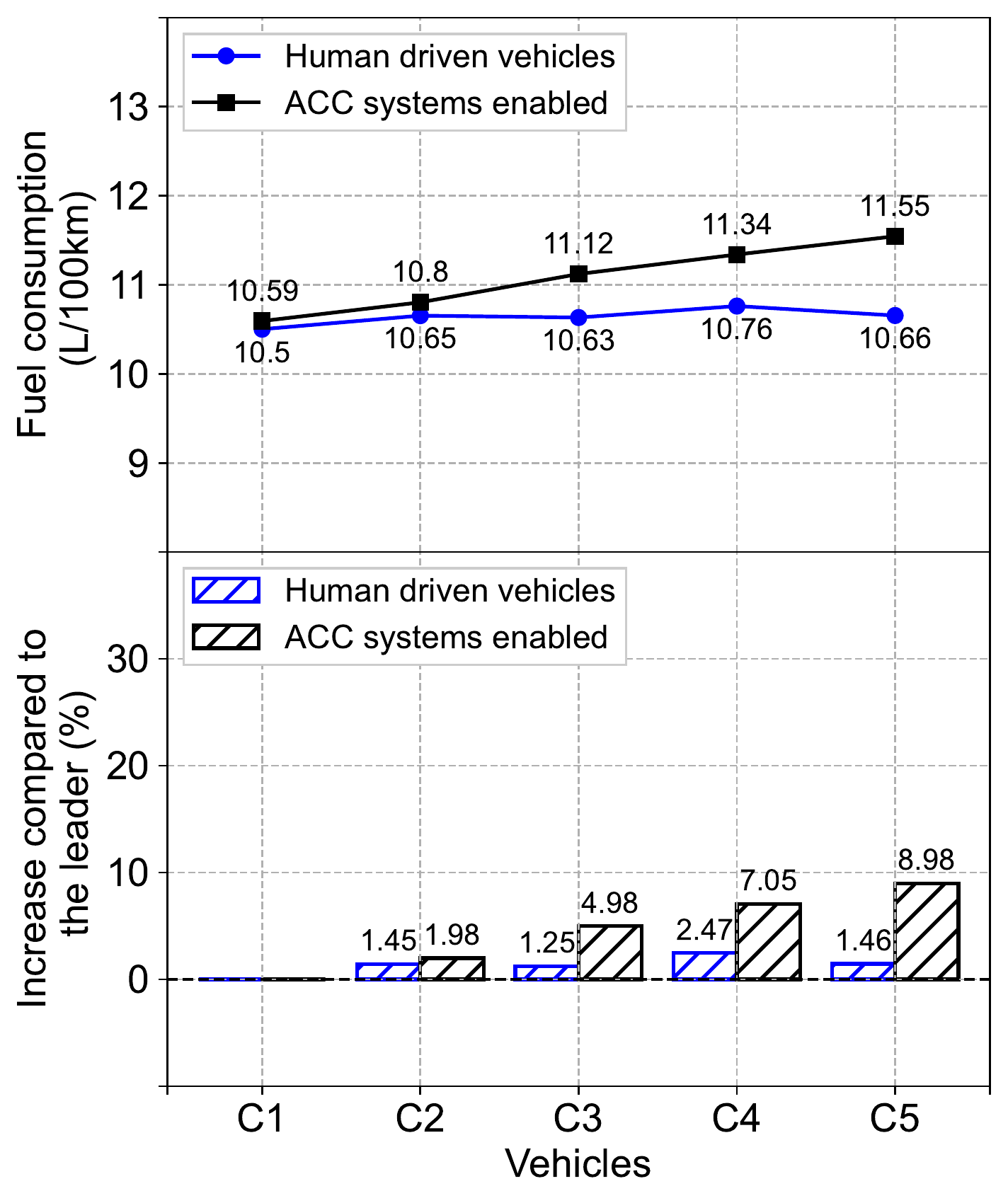}}
    \caption{Energy demand and fuel consumption for the AstaZero campaign.}
    \label{fig:Energy/Fuel_consumption_AstaZero}
\end{figure*}

Fig.~\ref{fig:Energy/Fuel_consumption_AstaZero} presents the absolute and relative energy estimation results per model for the AstaZero campaign. The tractive energy demand values in Fig.~\ref{fig:Tr_energy_AstaZero} are amplified upstream in the platoon with ACC engaged, primarily for the last participants (C3, C4, C5). On the other hand, human drivers show similar energy demand, with small variations in the estimated values. This is in agreement with the findings in the first campaign in Italy, confirming that stock ACC systems end up being ``energy-hungry".

Figs~\ref{fig:VT-micro_AstaZero}--\ref{fig:ARRB_AstaZero} display the fuel consumption per model for the same campaign. The platoon with ACC-enabled vehicles shows a clear upstream fuel consumption amplification. On the contrary, human counterparts realize very small value fluctuations, showing a steady-state behavior. As expected, the three fuel consumption models appear to differ again in absolute values. Concluding, the findings from both energy and fuel consumption models seem to agree with those mentioned above, confirming that commercially implemented ACC systems turn out to be less energy efficient compared to human drivers, from a platoon perspective.

\section{Further Analyses and Critical Discussion}\label{sec:discussion}

This section explores and discusses several aspects, including speed/acceleration profiles, string stability, road gradient, and standard deviation of speeds, behind the high energy footprint of ACC participants inside platoons. However, it reveals and highlights that commercially implemented ACC systems do not generally fail inside a platoon, in the sense that they deliver what is expected by their operational design. Their performance is conditioned to their functional specifications (spacing policy and its parameters) and environment. 

\subsection{What Affects Energy Consumption Most?}

\subsubsection{Speed and Acceleration Profiles}

\begin{figure*}[tbp]
     \centering
     \begin{tabular}{c}
     \subfloat[Human driven vehicles\label{fig:Speed/accel profiles_HDV}]{%
    \includegraphics[width=0.332\linewidth]{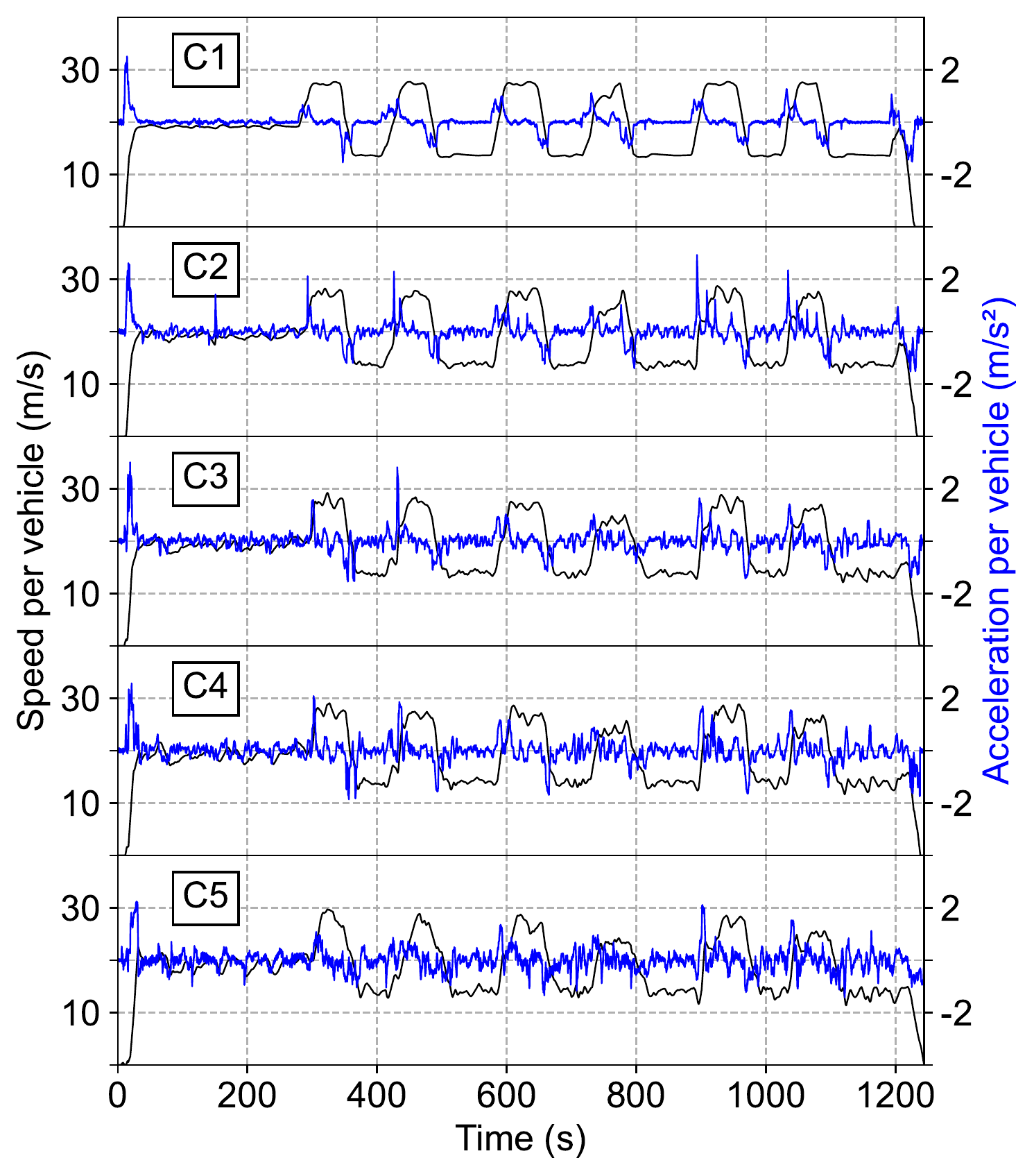}}
     \hfill
     \subfloat[ACC systems enabled\label{fig:Speed/accel profiles_ACC}]{%
    \includegraphics[width=0.332\linewidth]{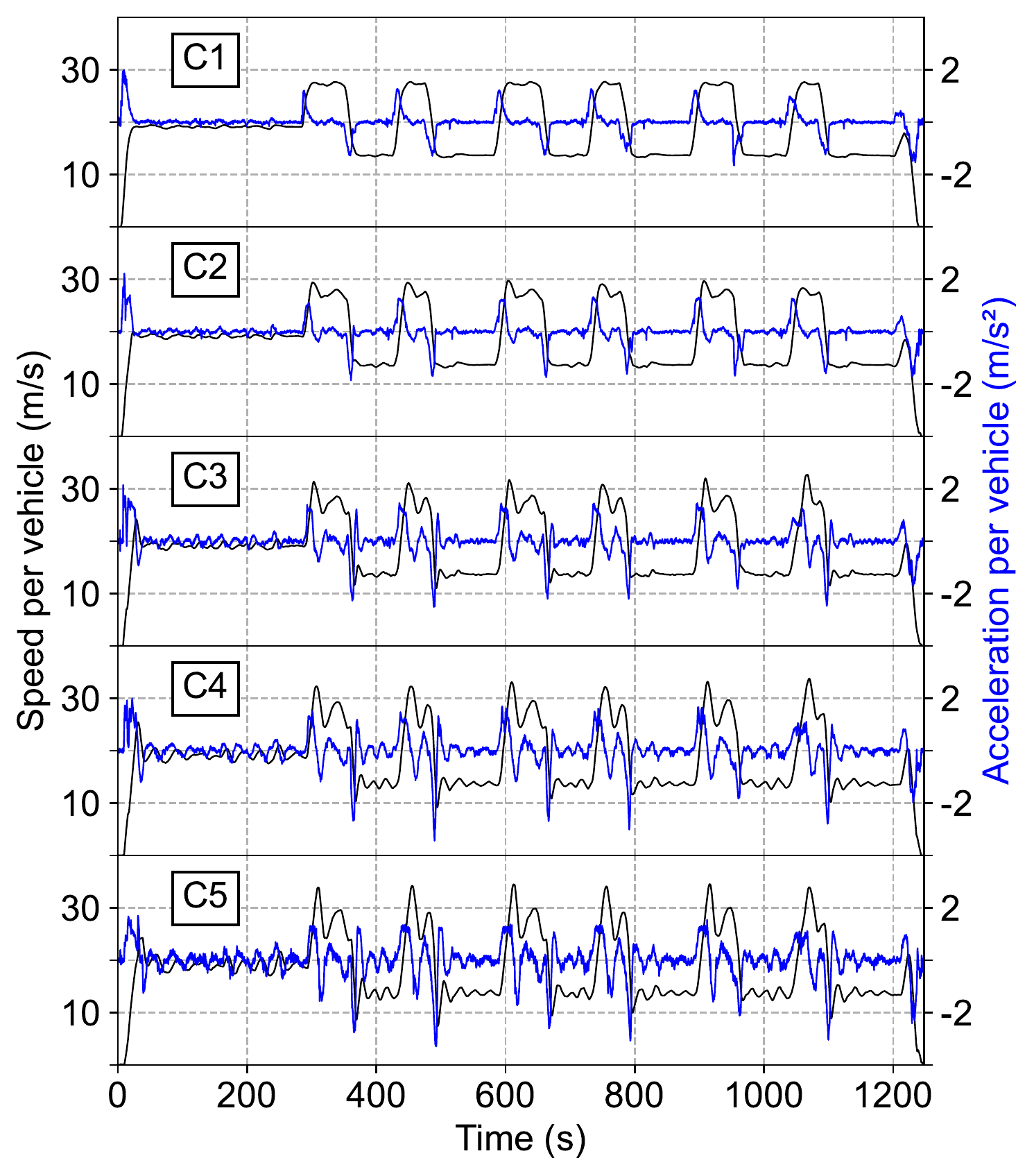}}
    \end{tabular}
    \hspace{-15pt} 
    \begin{tabular}{c}
        \subfloat[Steady parts\label{fig:Steady parts}]{%
         \includegraphics[width=0.3\linewidth]{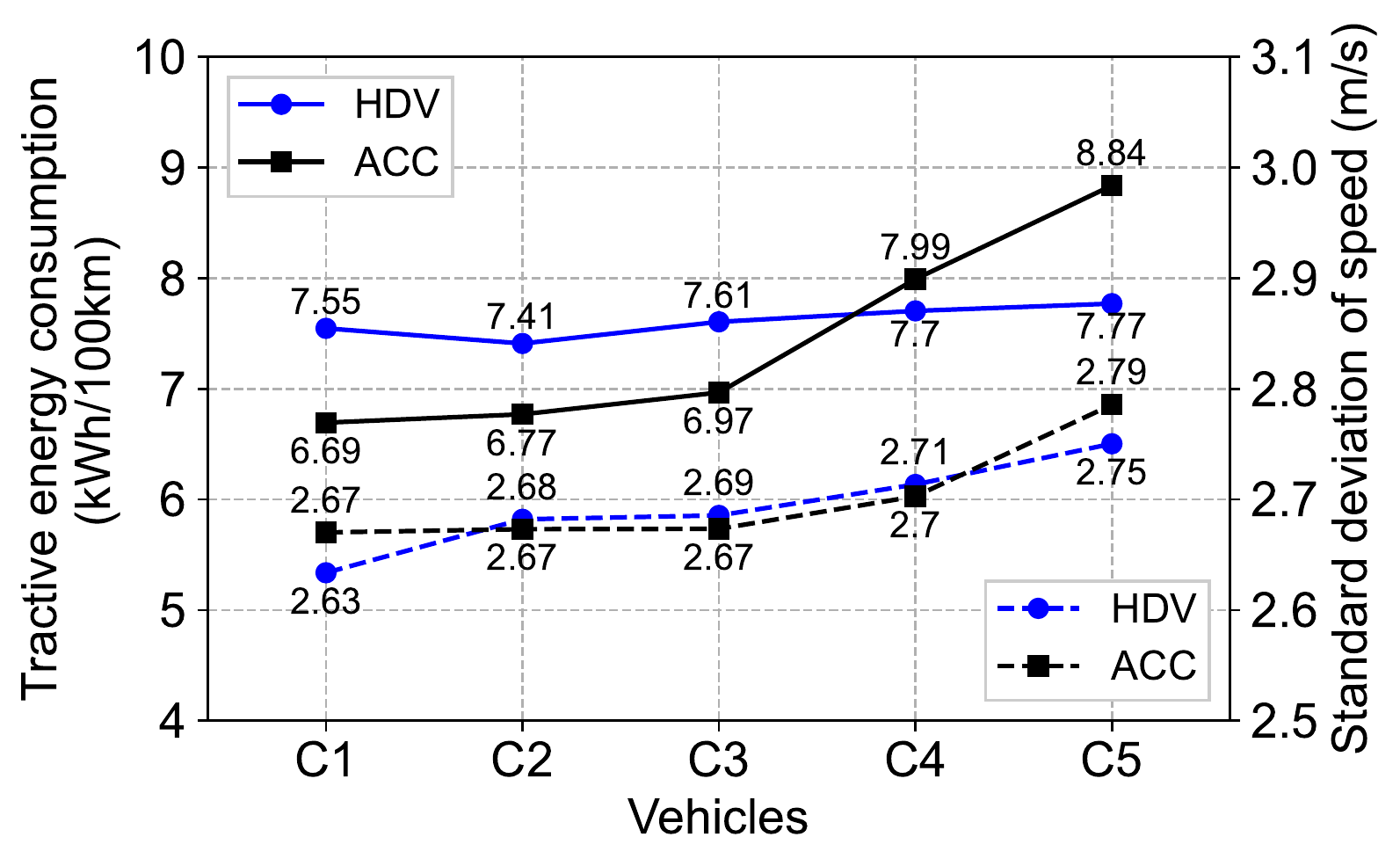}}\\ 
        \subfloat[Perturbation parts\label{fig:Perturbation parts}]{%
         \includegraphics[width=.3\linewidth]{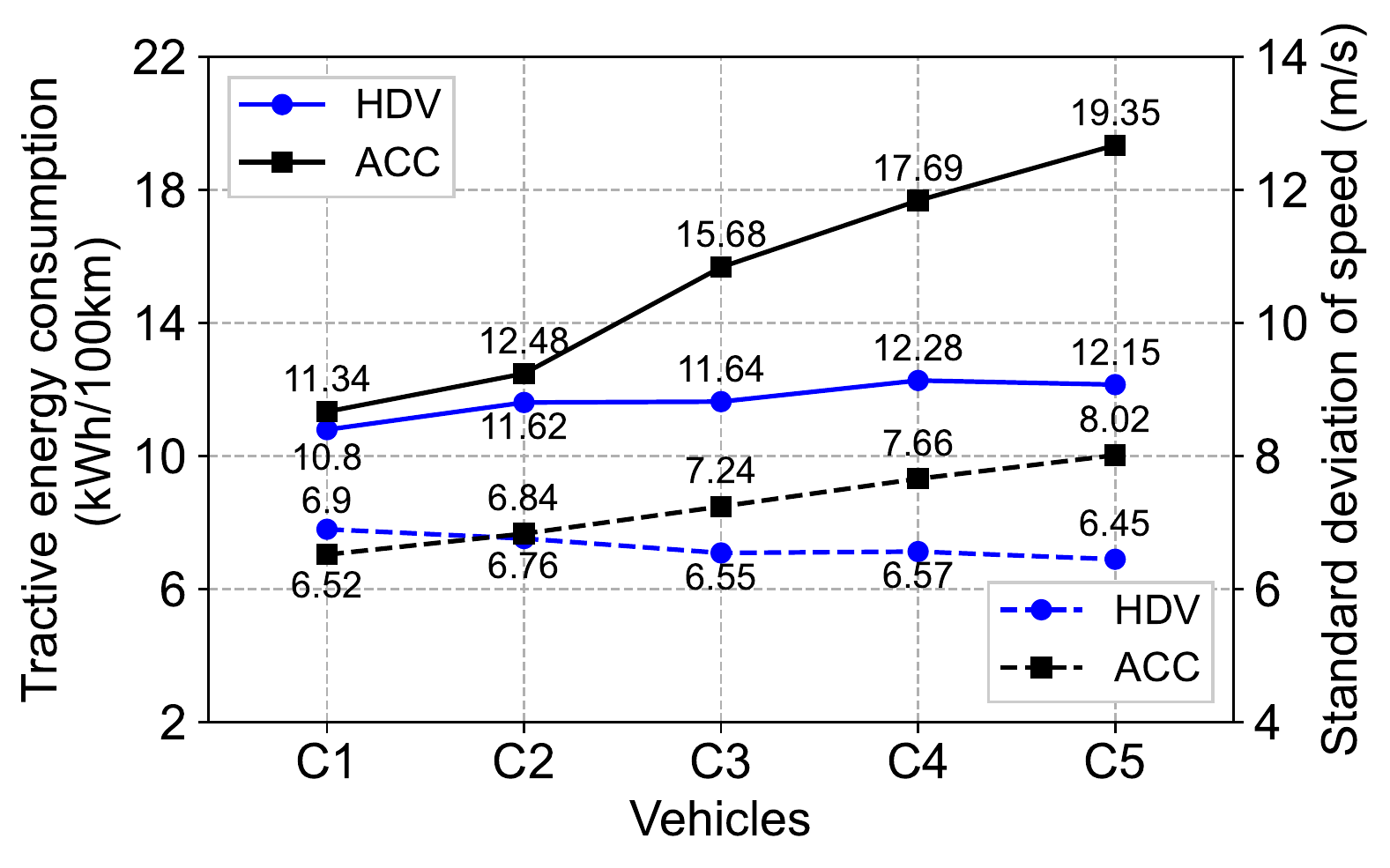}} 
    \end{tabular}
    \caption{Speed/acceleration profiles, tractive energy consumption, and standard deviation of speed  for the AstaZero campaign.}
    \label{fig:Speed/accel profiles}
\end{figure*}

By design, commercial ACC systems automatically adjust the vehicle’s speed by accelerating or decelerating it, to maintain a constant predefined time-headway with the vehicle in front or reach the desired preset speed. Figs~\ref{fig:Speed/accel profiles}a--\ref{fig:Speed/accel profiles}b illustrate the speed and acceleration profiles of each testing vehicle, separately, taken from the AstaZero campaign. Fig.~\ref{fig:Speed/accel profiles_ACC} shows the behavioral differences related to instantaneous speed and acceleration, between ACC systems and human drivers. The first couple of ACC systems in the platoon have low varied speed and acceleration values but the last vehicles in the platoon reveal strong and sharp accelerations with several peaks and variations in the speed profile. Human counterparts reveal smaller fluctuations around the target speed. Overall, the observed oscillations (peaks and spikes) in speed and acceleration of the commercially implemented ACC systems seem to affect the energy impact of the ACC participants inside the platoon. This is obvious since all models presented in Section \ref{sec:Methods} estimate energy demand and fuel consumption mainly, as functions of speed and acceleration.

Finally, it should be highlighted that the driving cycle itself plays an important role in both absolute and relative comparisons between modes. Specifically, frequent perturbations in a driving cycle (sharp accelerations/decelerations of the leader, changes in the speed limit, changes in the altitude, etc.) amplify the difference in the energy footprint between modes. In contrast, such differences might become negligible or even reversed during a normal highway driving cycle with minimal disturbances. 

\subsubsection{String Stability}

String stability refers to the mitigation of upstream propagating disturbances imposed by the leading vehicle inside the platoon \cite{8318387}. Under string stability, any non-zero disturbances in position, speed and acceleration of an individual vehicle in a platoon do not amplify upstream \cite{lei2012evaluation}. The following results from the AstaZero and Vicolungo campaigns show that ACC systems led to string unstable platoons \cite{doi:10.1177/0361198120911047}.

\begin{figure*}[tbp]
     \centering
     \subfloat[Human driven vehicles (AstaZero)\label{fig:Speed_oveshoot_HDV_Ast}]{%
         \includegraphics[width=0.329\linewidth]{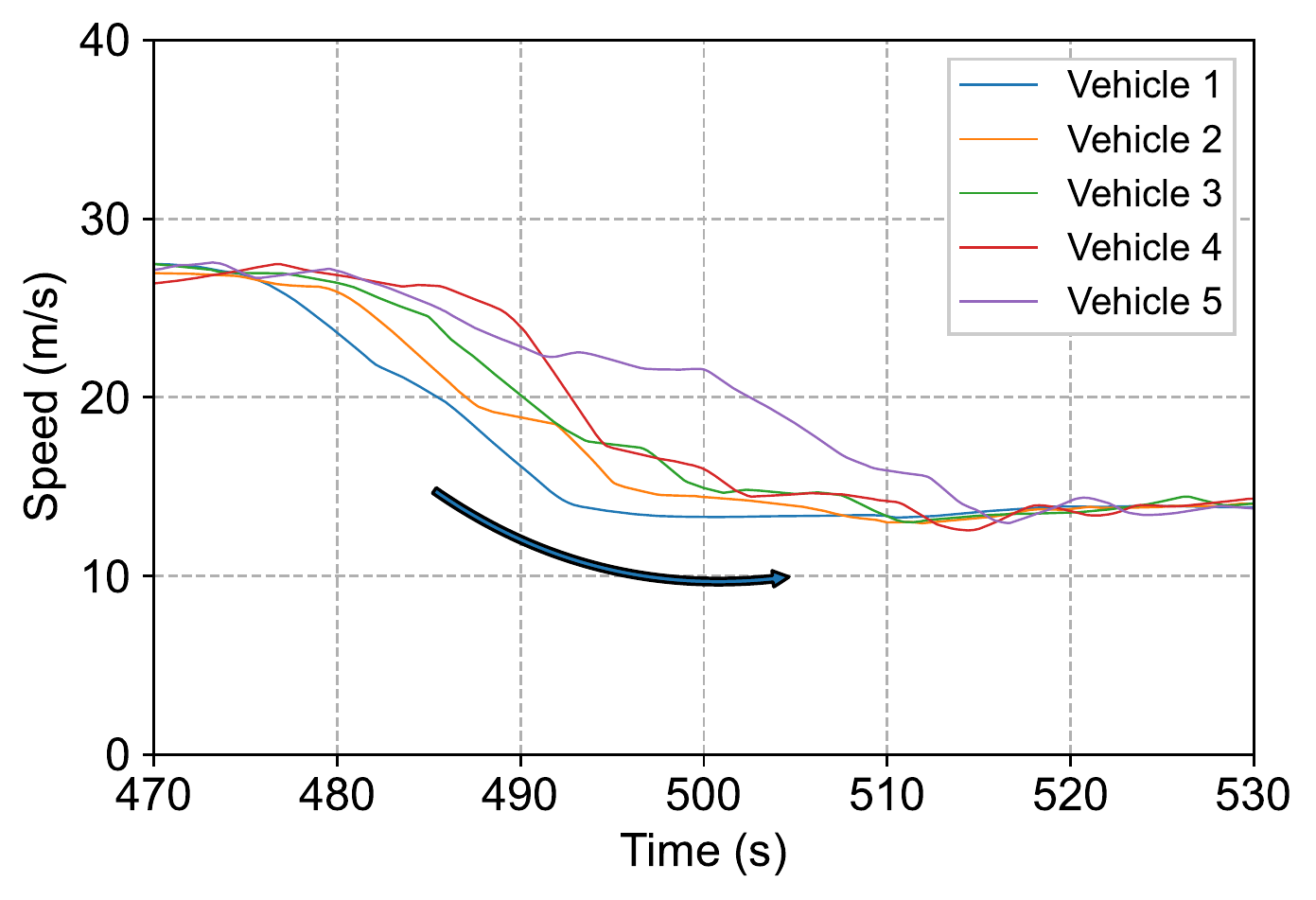}}
     \hfill
     \subfloat[ACC systems enabled (AstaZero)\label{fig:Speed_oveshoot_ACC_Ast}]{%
         \includegraphics[width=0.329\linewidth]{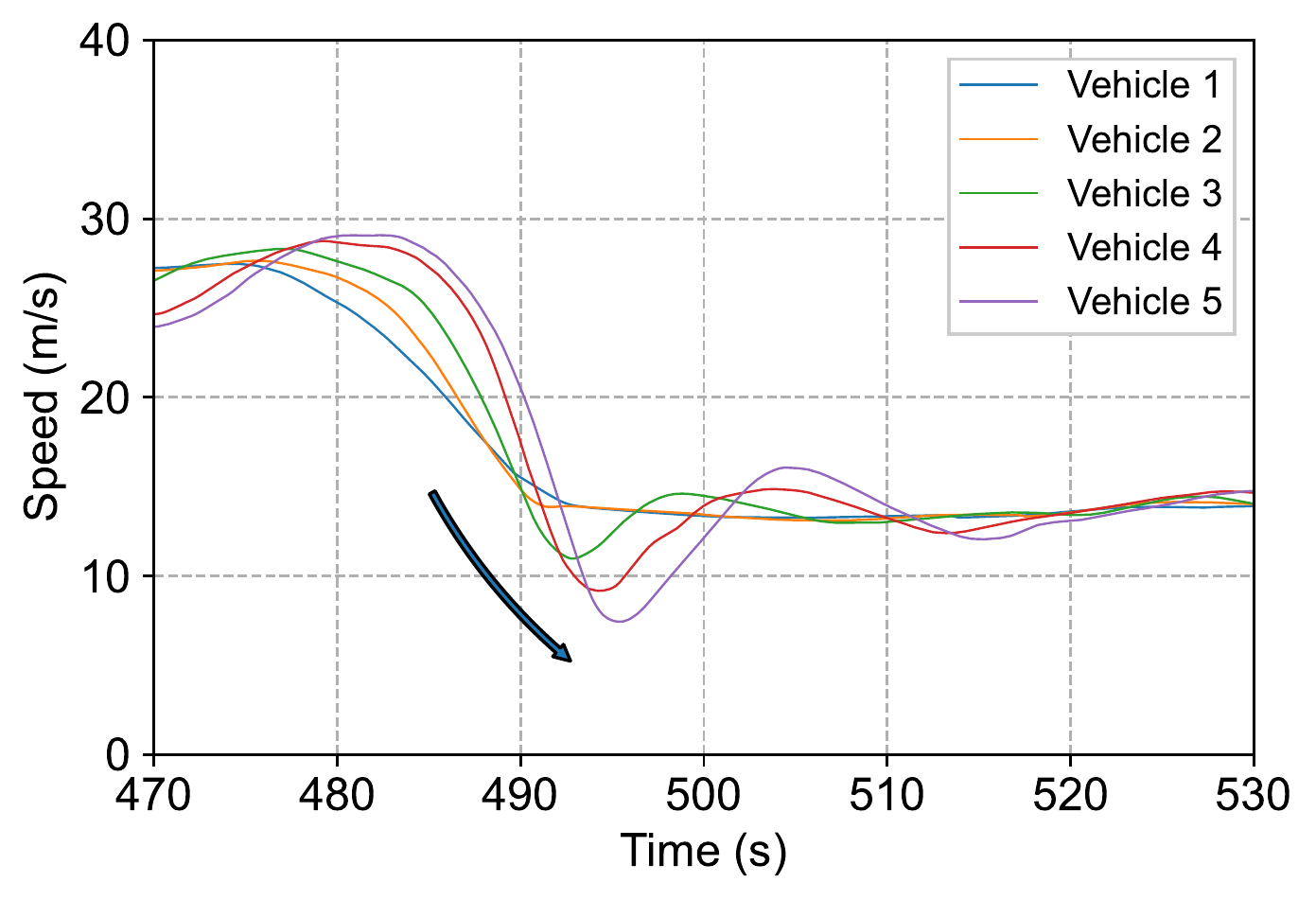}}
         \hfill
     \subfloat[ACC systems enabled (Vicolungo)\label{fig:Overshoots_It}]{%
         \includegraphics[width=0.329\linewidth]{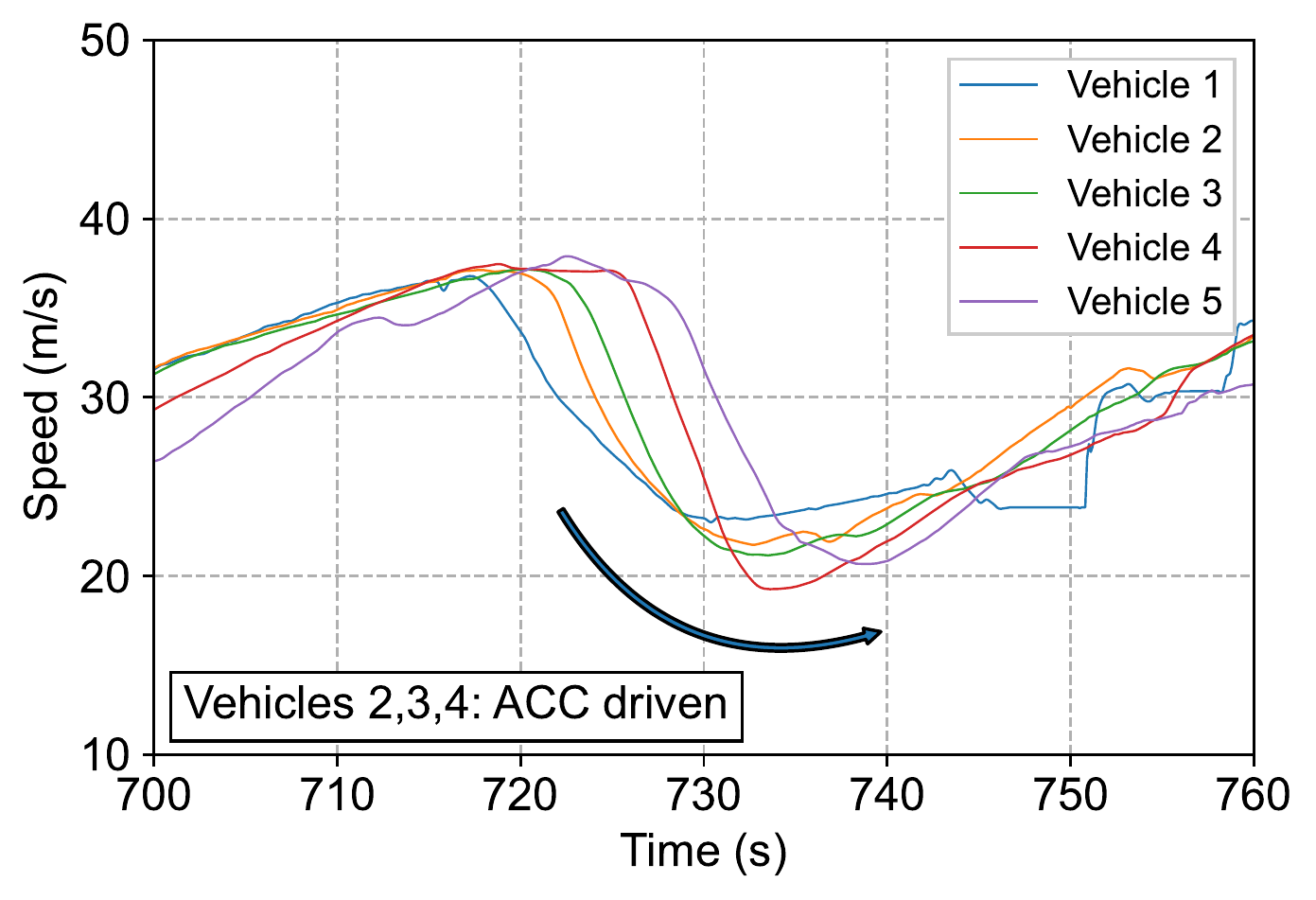}}
    \caption{Speed overshoots with human driven vehicles and ACC-engaged vehicles (AstaZero and Vicolungo).}
    \label{fig:Overshoots_Ast}
\end{figure*}

Figs~\ref{fig:Overshoots_Ast}a--\ref{fig:Overshoots_Ast}b depict two exemplary perturbation events, which occurred between the same speeds for both driving modes. The ACC drivers reveal several peaks and steep variations in the speed and acceleration profiles, especially when moving upstream the platoon, see  Fig.~\ref{fig:Speed/accel profiles}b. In Fig.~\ref{fig:Overshoots_Ast}a, the speed variations of the leader are not amplified through the human drivers. However, in Fig.~\ref{fig:Overshoots_Ast}b, ACC followers significantly enlarge their leader’s speed perturbation, revealing large speed overshoots amplified upstream to the platoon. The black curved arrows in Figs~\ref{fig:Overshoots_Ast}a--\ref{fig:Overshoots_Ast}b depict the stable car-following behavior of human drivers and the string instability of the ACC participants, respectively.

String instability of ACC systems negatively affects their energy impact. Analytically, string stability is studied (among others) with the $\mathcal{L}_2$ norm, which in turn is directly related to the square of the speed or energy. A correlation between string stability and energy and fuel consumption seems to exist also in real-world observations, with speed overshoots propagating upstream the platoon directly affecting ACC's energy and fuel consumption. Again, this can be attributed to the fact that instability leads to higher acceleration and speed variation, the key ingredients for elevated fuel consumption estimates by the models. Systematic analysis of this relation should be performed across the full operational domain (driving cycles) and under many possible conditions (road and routes), to obtain meaningful and solid results.

The human participant (C5) in the NB route of the first campaign in Italy, revealed a relatively downward value trend among the various fuel and energy consumption models, trying to absorb the energy propagation that occurred upstream of the platoon. Fig.~\ref{fig:Overshoots_Ast}c depicts one perturbation event during real-world conditions in the motorway of Autostrada A26 in Italy. As can be seen, the last follower (C5) absorbs the speed overshoots propagating upstream of the platoon. A possible explanation could be that human drivers can anticipate decelerations occurring two to three vehicles downstream of the platoon, and hence respond in fast. Therefore, this tendency for absorbing the perturbations propagating upstream of the platoon could be related to the lower energy and fuel consumption values achieved by the last human participant.

\subsubsection{Road Gradient}

Road gradient is a source of perturbation for vehicles since it poses additional forces.  As such, it is closely related to sting instability and energy consumption. The test section of the first experimental campaign in Italy contains various uphill and downhill segments with lots of steep points. The road gradient in these segments is an additional input to the ACC controllers. Under steady-state conditions, an altitude increase would require additional effort from an ACC-engaged vehicle to reach the pre-specified reference speed or time-gap. In this effort, trying to counterbalance the impact of the road gradient, the ACC system ends up overshooting. In a similar way, as the altitude decreases, the ACC system undershoots reaching a lower speed than the desired one. In both cases, oscillations are generated around the desired speed (of the leader) directly affecting the followers. These oscillations are propagating upstream of the platoon and could lead to string instability and hence, affect the energy impact of ACC systems. According to \cite{doi:10.1177/0361198120911047}, even for slight perturbations derived by variability in the road gradient, string instability can be observed, raising concerns about potential consequences in traffic flow as the penetration rate of ACC systems is increasing.

\subsubsection{Standard Deviation of Speeds}

Figs~\ref{fig:Speed/accel profiles}c--\ref{fig:Speed/accel profiles}d show the tractive energy consumption values in relation to those of standard deviation of speed, for both steady (see Fig.~\ref{fig:Speed/accel profiles}c) and perturbation (see Fig.~\ref{fig:Speed/accel profiles}d) parts, for the second experimental campaign in AstaZero. Steady parts are defined as those where the drivers maintain a constant speed with very small fluctuations around the equilibrium points applied by the leader in the platoon (see Figs~\ref{fig:Speed/accel profiles}a--\ref{fig:Speed/accel profiles}b), while perturbation parts are those where the leader performs scheduled deceleration or acceleration events. 

As can be seen in Figs~\ref{fig:Speed/accel profiles}c--\ref{fig:Speed/accel profiles}d, there is an agreement between the tendency both set of values reveal (tractive energy consumption versus standard deviation of speeds), developing a cause and effect relationship. This statement naturally derives from the fact that the input both drivers “see”, is the term of speed. At the same time, tractive energy consumption is mainly estimated as a function of speed and acceleration. Therefore, the standard deviation of speed is an appropriate parameter to explain and quantify the changes in tractive energy consumption values. 

For the steady parts (see Fig.~\ref{fig:Speed/accel profiles}c), a somehow conservative behavior is revealed for both human and ACC drivers, with the exception of the two last followers in the ACC platoon due to small, contingent perturbation events. On the contrary for the perturbation events (see Fig.~\ref{fig:Speed/accel profiles}d) there is a clear amplification in the ACC platoon, while a noticeably invariable behavior in the human platoon, regarding both sets of values. 

\subsection{The Merits of Commercially Implemented ACC Systems} 

Despite the discussed shortcomings of ACC systems, this section highlights that commercially implemented ACC systems do not generally collapse inside a platoon.

\subsubsection{Space and Time-Gap Distributions}

\begin{figure*}[tbp]
     \centering
     \subfloat[Time-gap (Vicolungo)\label{fig:Time-gap_It}]{%
         \includegraphics[width=0.249\linewidth]{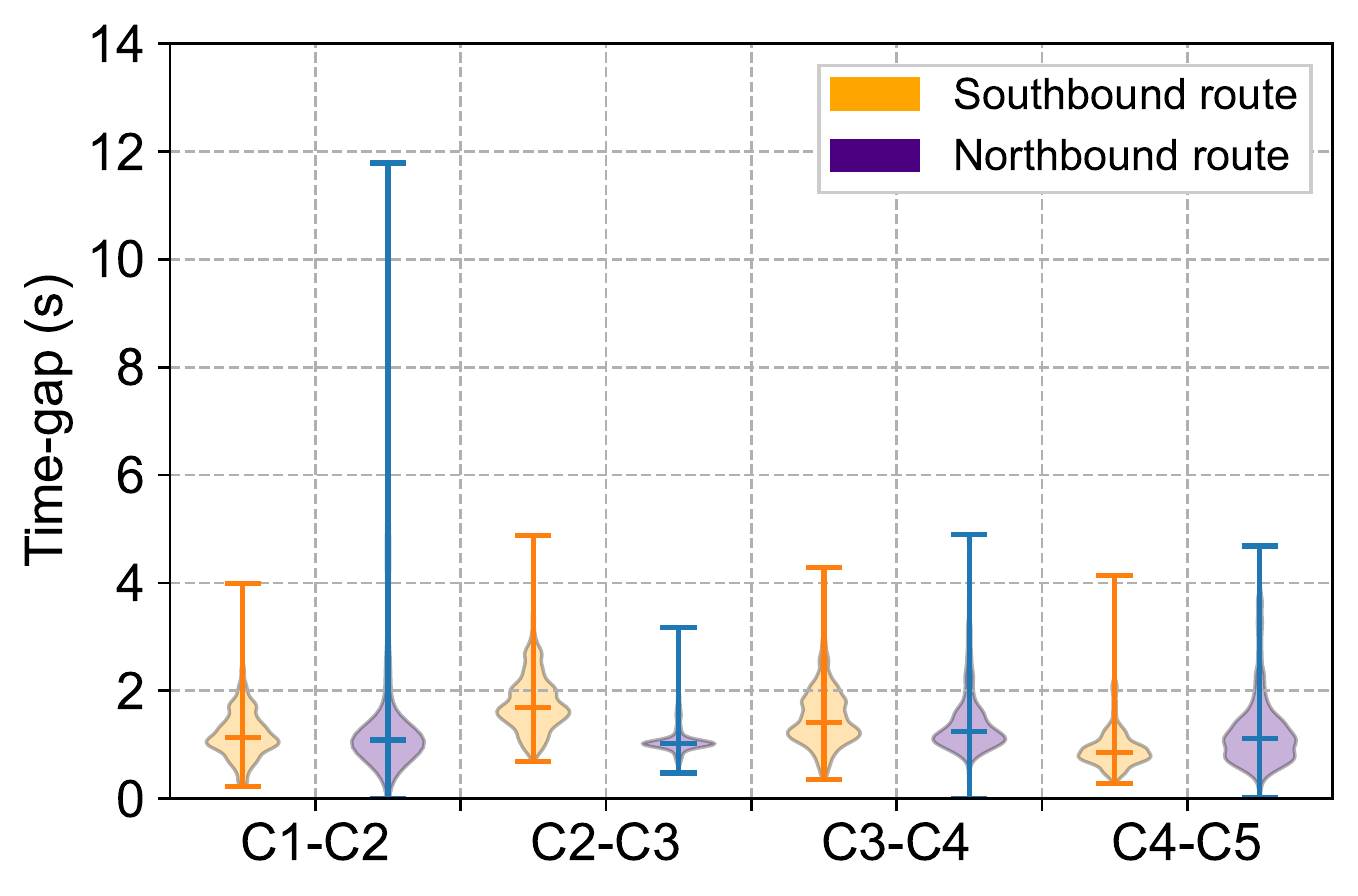}} 
     \hfill
     \subfloat[Space-gap (Vicolungo)\label{fig:Space-gap_It}]{%
         \includegraphics[width=0.253\linewidth]{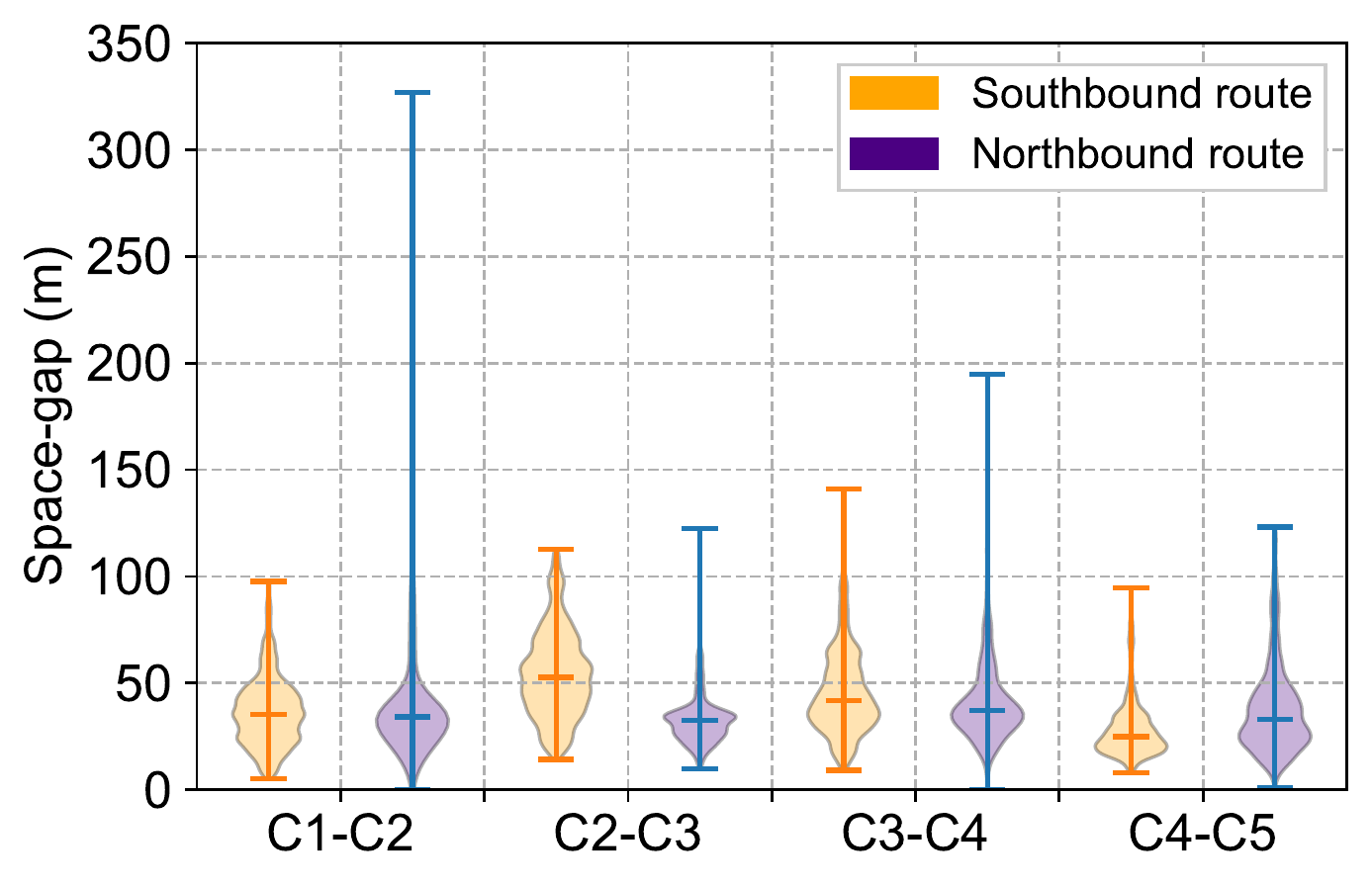}} 
         \hfill
\subfloat[Time-gap (AstaZero)\label{fig:Time-gap_Ast}]{%
         \includegraphics[width=0.244\linewidth]{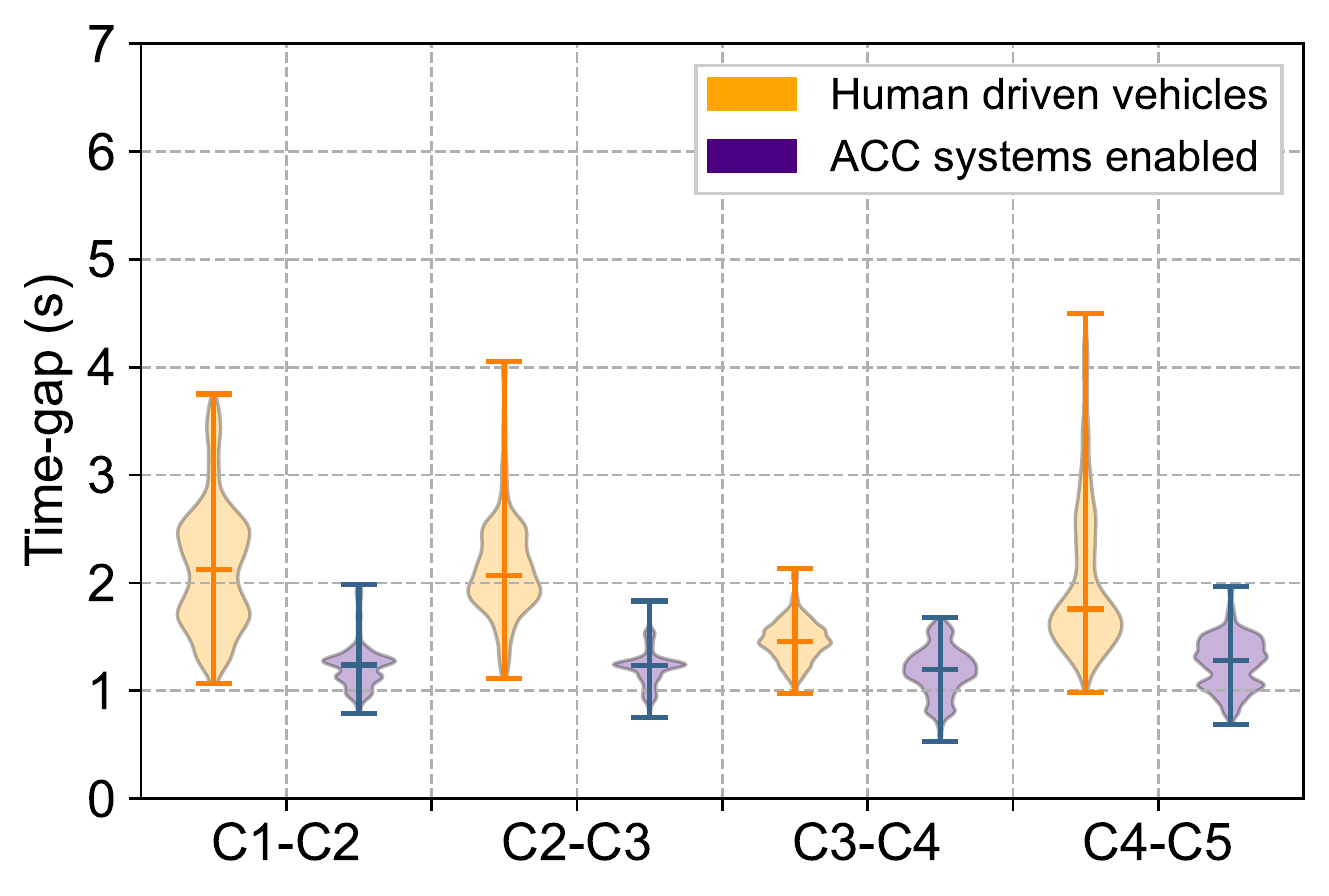}} 
     \hfill
     \subfloat[Space-gap (AstaZero)\label{fig:Space-gap_Ast}]{%
         \includegraphics[width=0.252\linewidth]{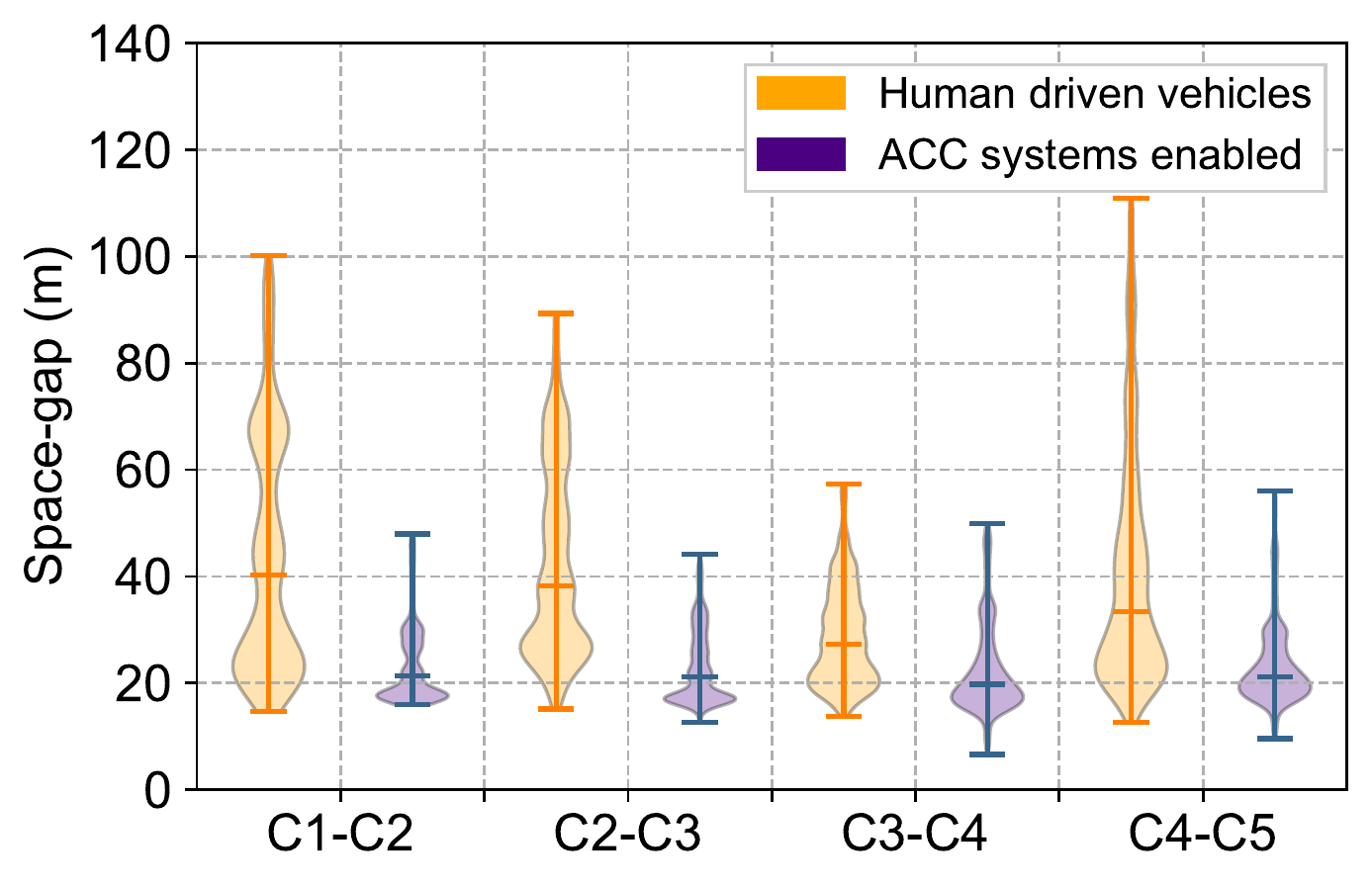}}
    \caption{Time and space-gap distributions for the two campaigns.}
    \label{fig:Space-Time_gaps}
\end{figure*}

The logic behind ACC systems design focuses on its functional specifications (spacing policy and its parameters) rather than eco-friendly driving instructions. As a consequence, ACC systems outperform human-driven vehicles in these areas. Fig.~\ref{fig:Space-Time_gaps} displays the time-gap and space-gap distributions for both experimental campaigns held in Italy and Sweden. The computation of time-gap between two vehicles inside a platoon was performed by dividing the obtained inter-vehicular spacing (IVS) between two vehicles by the speed of the proceeding vehicle. Space-gap was already estimated (via the IVS) and hence, available in the two data sets. As can be seen, the time and space-gap values with ACC systems engaged are smaller (in absolute values), more steady, and better distributed compared to those of human drivers. 

This favorable result for the ACC drivers is attributed to the stock ACC controller of the vehicles involved in the trials. It highlights that commercially implemented ACC systems, despite the criticism of their negative impact in relation to traffic flow, stability, and energy demand, deliver some quantified improvements.  Actually, ACC systems deliver the improvements that are expected from their design and controller synthesis e.g.\ to adhere to a constant time-headway policy (the so-called CTHP). However, keeping a constant headway policy comes with a price. It is well known that constant space-headway policies (CSHP) lead to string unstable platoons of vehicles \cite{Seiler, Jovanovic, Middleton}, while constant time-headway policies (CTHP) have been shown to be string stable for ACC vehicles without inter-vehicle connectivity \cite{Ioannou_Chen, liang_stability, Swaroop_stability, Rajamani}. In this study, provided that the time-gaps are more or less constant with ACC systems engaged, our conjecture is that the commercially implemented ACC systems of the involved vehicles in the trials employ a constant time-headway policy. However, the core functionality and architecture (controller type and its parameters) are not publicly available, so any conclusions must be drawn with caution.

\subsubsection{Driving Environment}

Another remarkable observation is related to the comparison of the same graphs in Fig.~\ref{fig:Space-Time_gaps} for the two different campaigns, one took place in a public motorway stretch and the other in a protected rural environment. In the protected environment of AstaZero, it seems the improvements for the space and time-gaps are more evident and higher compared to those in Italy (cf.\ Figs.~\ref{fig:Space-Time_gaps}(a)--\ref{fig:Space-Time_gaps}(b) with Figs.~\ref{fig:Space-Time_gaps}(c)--\ref{fig:Space-Time_gaps}(d)). This might be attributed to: (a) the employed vehicles (their characteristics and specifications); (b) the environment (protected in AstaZero, so no inference with other cars and mixed heterogeneous traffic; public motorway in Italy with mixed traffic and disturbances). These observations also underline that stock ACC systems in protected environments and experimental campaigns might perform better but can be ideally tested in adverse environments.

\begin{figure*}[tbp]
     \centering
     \subfloat[Human driven vehicles\label{fig:Time-gap_Speed_AstaZero_scatter_HDV}]{%
         \includegraphics[width=.49\linewidth]{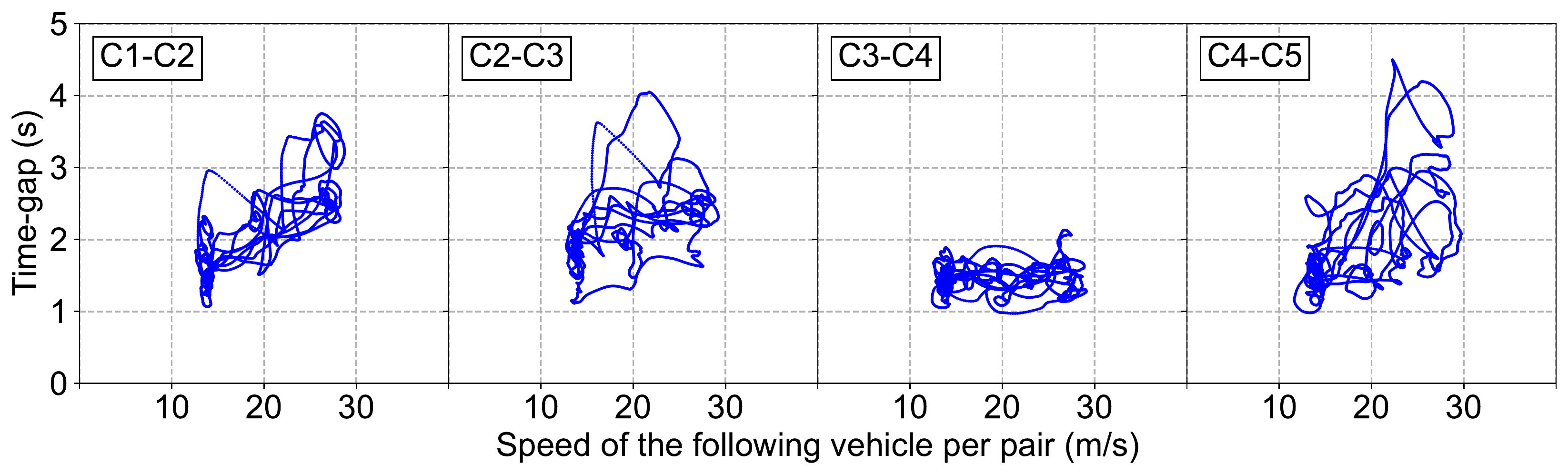}} 
    \hfill
     \subfloat[Human driven vehicles\label{fig:Space-gap_Speed_AstaZero_scatter_HDV}]{%
         \includegraphics[width=.5\linewidth]{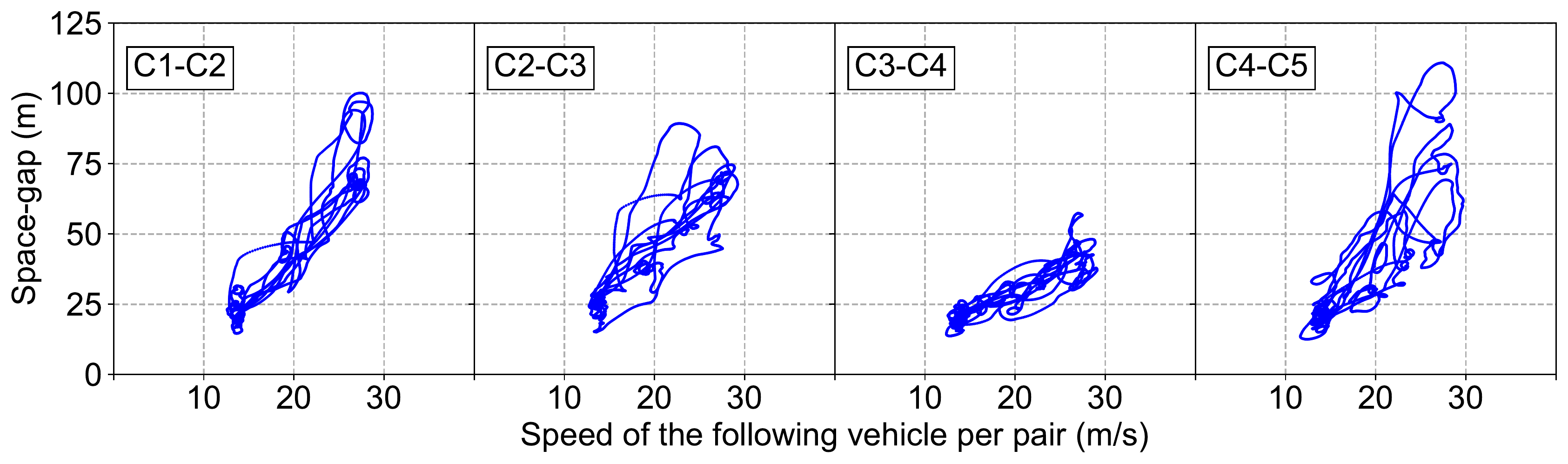}}\\ 
     \subfloat[ACC-engaged ego vehicles\label{fig:Time-gap_Speed_AstaZero_scatter_ACC}]{%
         \includegraphics[width=.49\linewidth]{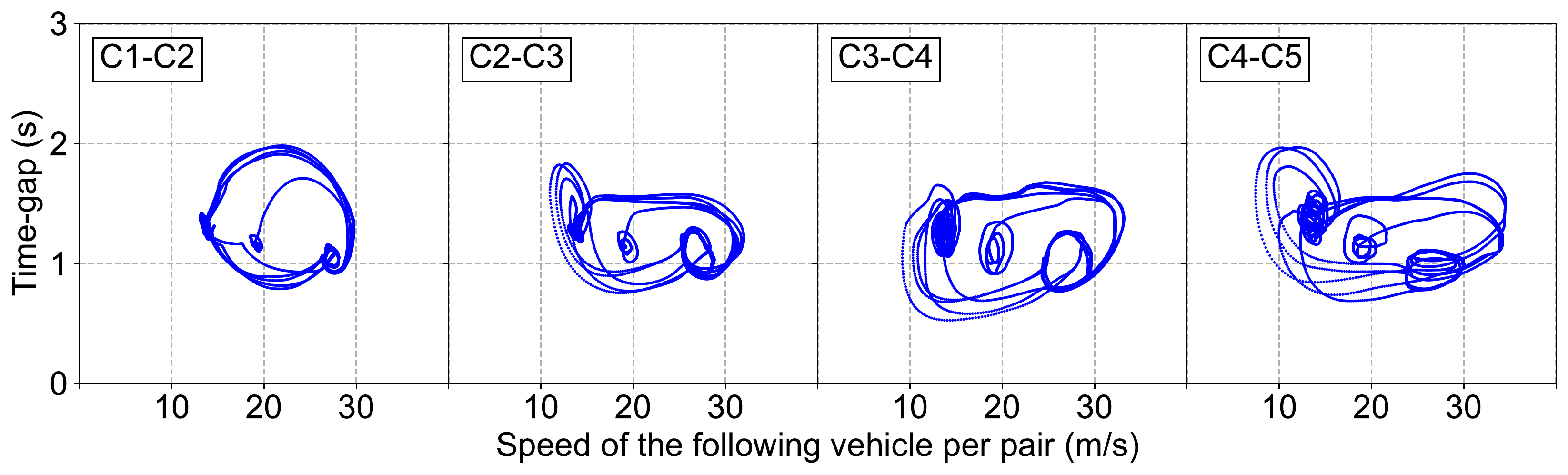}} 
\hfill
\subfloat[ACC-engaged ego vehicles\label{fig:Space-gap_Speed_AstaZero_scatter_ACC}]{%
         \includegraphics[width=.5\linewidth]{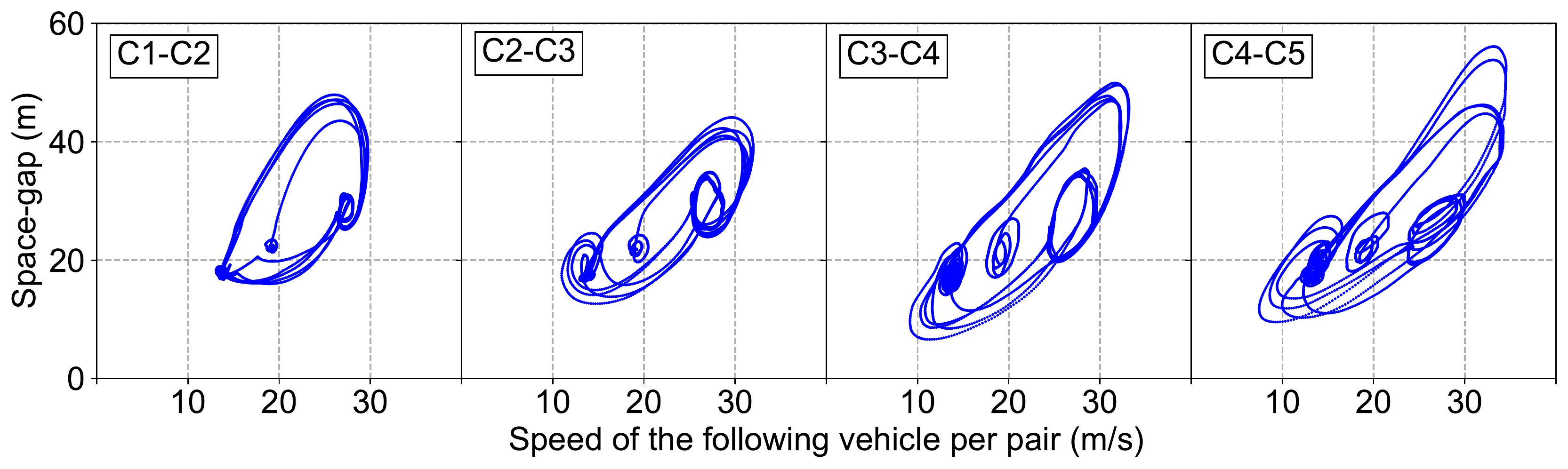}} 
    \caption{Time and space-gap scatter plots for the AstaZero campaign.}
    \label{fig:Scatter_plots_AstaZero}
\end{figure*}

\subsubsection{ACC System Specifications and Spacing Policy}

A further analysis on the time and space-gaps is presented in Fig.~\ref{fig:Scatter_plots_AstaZero} for the AstaZero campaign. As can be seen in Figs~\ref{fig:Scatter_plots_AstaZero}a--\ref{fig:Scatter_plots_AstaZero}b, human drivers keep larger distances as speed increases with a somewhat linear correlation, indicating that human drivers implement a proportional policy (P-type controller in control theory) to compensate in response to traffic perturbations. This somehow explains the argument that people use less power in the vehicle (in comparison to its potential) at higher speeds since human drivers behave linearly.

However, the ACC drivers in Figs~\ref{fig:Scatter_plots_AstaZero}c--\ref{fig:Scatter_plots_AstaZero}d, as expected, keep constant distances for different speeds, i.e., a constant time-headway policy. The ACC driving behavior indicates a periodic cyclic operation, as trajectories contained in ``limit cycles". This is a favorable behaviour and relevant to ACC stability both individually and as a group (i.e.\ string stability) since all trajectories in the interior of a limit cycle approach the limit cycle for time approaching infinity. This behaviour is conjectured to be attributed to the Proportional-Integral-Derivative (PID-type controller) constant time-headway policy  employed in most of the commercial vehicles available in the market \cite{MILANES2014}. However, as already noted, the core functionality and architecture of stock ACC systems (controller type and its parameters) are not publicly available, so any conclusions must be drawn with caution. As a final remark, it is clear from the figures that the two drivers (humans and ACC) are operating in different phase-space domains, thus there is room for improvement for both agents.

\begin{figure*}[tbp]
     \centering
     \subfloat[Human driven vehicles\label{fig:Contour_plot_HDV_AstaZero}]{%
         \includegraphics[width=.7\linewidth]{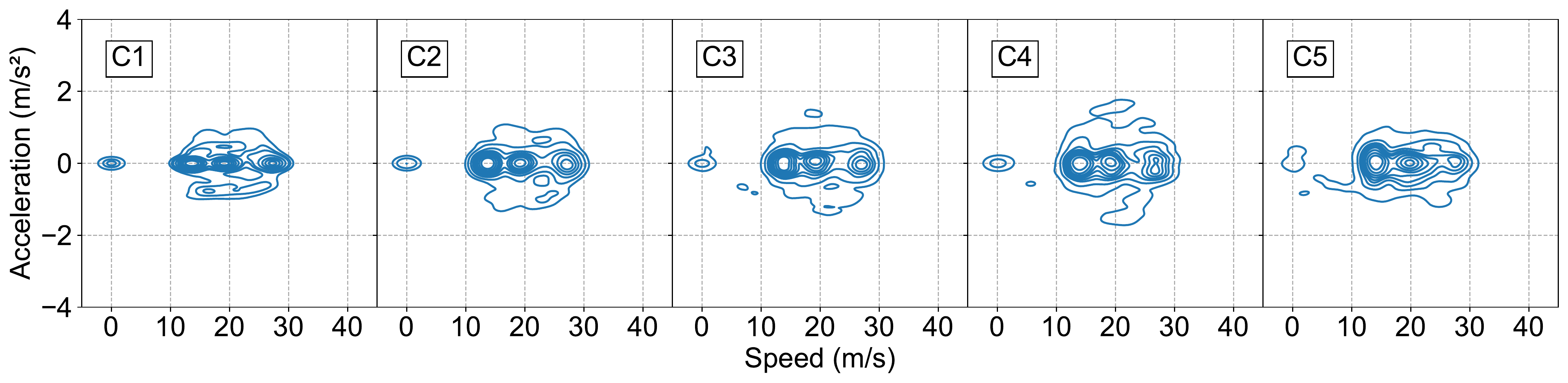}}\\
     \subfloat[ACC-engaged vehicles\label{fig:Contour_plot_ACC_AstaZero}]{%
         \includegraphics[width=.7\linewidth]{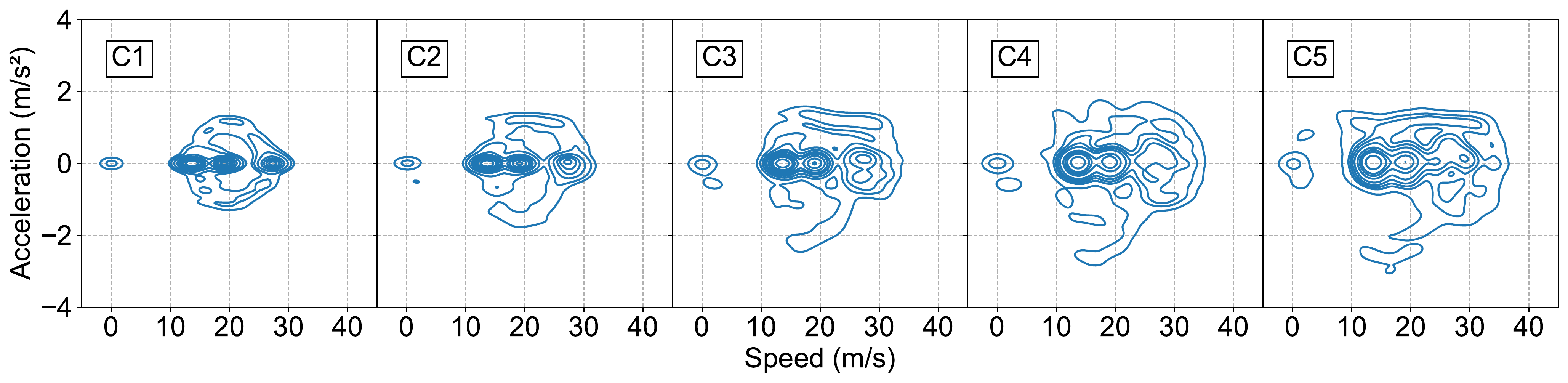}}
    \caption{Speed-acceleration joint probability distributions for the AstaZero campaign.}
    \label{fig:Contour_plot_AstaZero}
\end{figure*}

\subsubsection{The Bond Between ACC-engaged Ego Vehicles and Human Drivers}

Fig.~\ref{fig:Contour_plot_AstaZero} depicts the speed-acceleration joint probability distributions for the AstaZero campaign. These contour maps can be used to further study the correlation between the leader’s and the follower’s speed-acceleration driving profile \cite{article}. Such analysis would be useful to unveil how strong the bond between the (free) leader and the ego vehicles in the platoon is. Humans have the tendency to follow the driving pattern of the leader, while ACC drivers generally do not observe the leader; they use the leader's speed as input for control of their speed. From a first inspection, it is clear that both participants indicate quite similar contour plots inside their platoons, with ACC counterparts producing more scattered plots than human drivers do due to higher deceleration values, especially in perturbation events. Also, in Fig.~\ref{fig:Contour_plot_AstaZero}(b) ACC's counterplots tend to be more and more scattered as we move upstream the platoon. This result is in total agreement with the findings mentioned above regarding speed overshoots being amplified upstream of the ACC platoon, causing string instabilities. 

To further illustrate this, we quantify the similarity between the distribution maps of the participants with respect to their leader (i.e., correlation of the pairs C1-C2, C1-C3, C1-C4, and C1-C5) using the Pearson correlation coefficient. As can be seen in Table \ref{tab:cor_coef_AstaZero}, in both driving modes there is a tendency for upstream bond suppression inside the platoon. However, the obtained correlation values are more favorable for the ACC participants indicating a very strong bond between the ACC participants and their leader. As it turns out, the tendency of human drivers to reproduce the driving patterns of the vehicles in front did not play a decisive role in the results. However, a closer look in Fig.~\ref{fig:Speed/accel profiles}(b) could simplify the obtained results; since the speed profiles of the ACC participants are quite the same, with a small time-shift and some maximization in their values (overshoots) as we move upstream the platoon, especially during perturbation events.

In conclusion, ACC systems do not generally fail inside a platoon. They turn out to be more effective in terms of functional specifications attributed to ACC design and synthesis. Hence, the fact that manufacturers set different priorities than expected regarding environmental concerns is perfectly reasonable. But everything comes with a cost: the more efficient ACC systems are regarding their functional specifications, the less energy efficient tend to be. 

\begin{table}[tbp]
\small
     \caption{Correlation Coefficients (AstaZero Test Track).}\label{tab:cor_coef_AstaZero}
     \centering
     \begin{tabular}{c|cccc}
              \hline
              & C1--C2 & C1--C3 & C1--C4 & C1--C5 \\
              \hline
              Human Driven Vehicles & 0.77 & 0.64 & 0.54 & 0.44 \\
              ACC-engaged Ego Vehs & 0.92 & 0.88 & 0.70 & 0.61 \\ 
              \hline
          \end{tabular}
\end{table}

\section{Conclusions and Outlook}\label{sec:Conclusion}

This study assessed the energy impact of commercially implemented ACC systems under car-following formation. The tractive energy demand and fuel consumption for human-driven and ACC-engaged vehicles in real-life experimental campaigns were assessed, with a variety of vehicle specifications, propulsion systems, and road and traffic conditions. To this end, high-resolution empirical data from two experimental car-following campaigns were used. Then, energy demand and fuel consumption estimations were calculated by employing four independent models. Finally, the study was expanded to capture the behavioral similarities and differences between the considered  driving modes for a deeper and better understanding of stock ACC energy footprint and its causes. This includes ACC operation (e.g.\ strong accelerations) and its impact inside the platoon (e.g.\ string instabilities), the relation of the latter to ACC energy demand, and ACC functional specifications (e.g.\ time and space-headways).

The main findings of the present paper are as follows:
\begin{itemize}
    \item Commercially implemented ACC systems are less energy efficient, revealing a tendency for upstream energy propagation inside platoons.
    \item Human counterparts adopt a more conservative and steady energy behaviour. 
    \item ACC driving operation and behaviour (strong accelerations, steep speeds, etc.) may negatively affect the energy impact of stock ACC systems under car-following conditions.
    \item Commercially implemented ACC systems may lead to string instability failing to avoid an upstream energy amplification.
    \item Commercially implemented ACC systems succeed in delivering a constant time-headway policy. 
    \item Commercially implemented ACC systems develop a very strong bond with their leader inside platoons.
    \item ACC driving indicates a periodic cyclic operation, as time/space-gap vs.\ speed trajectories contained in ``limit cycles”. 
    \item Human drivers and stock ACC systems are operating in different time/space-gap vs.\ speed  domains. Thus there is room for improvement for both agents.
\end{itemize}

It should be highlighted that the more efficient stock ACC systems are regarding their functional requirements and architecture, the less energy efficient they tend to be. For instance, the ACC participants succeeded in maintaining a predefined constant  headway policy, which is part of the stock ACC system logic and one of the requirements vehicle manufacturers need to fulfill, with both time and space-gap distributions accumulated around lower levels compared to human ones, indicating that commercially implemented ACC systems are more efficient. On the contrary, stock ACC systems failed to maintain a low energy profile compared to human drivers, ending up being less energy efficient. Consequently, commercially implemented ACC systems must be designed to achieve a trade-off between functional requirements and tractive energy and fuel consumption (eco-driving instructions). This will ideally eliminate high energy levels and achieve the desired safety features of commercial ACC systems. 

Despite the criticism of commercial ACC systems for a negative impact on energy demand, fuel consumption, and stability, they seem to operate reliably and efficiently under steady-state conditions and they are likely to improve in the near future. Future work will focus on possible solutions for the instability and energy inefficiency of ACC systems, including:
\begin{itemize}
    \item Multianticipation, where an ACC driver will be able to observe the driving behaviour of other vehicles downstream of the platoon, similarly to human drivers. This is already possible in some Cooperative ACC (CACC) systems, see e.g. \cite{DONA2022103687}.
    \item Enhanced connectivity with V2X (Vehicle-to-Vehicle and Vehicle-to-Infrastructure) communication, to ensure safe and quick response to perturbation events much further downstream in a timely manner, and thus, generating smoother responses, see e.g. \cite{9758647}. 
\end{itemize}

\bibliographystyle{IEEEtran}
\bibliography{IEEEabrv,Bibliography}

\begin{IEEEbiography}[{\includegraphics[width=1in,height=1.25in,clip,keepaspectratio]{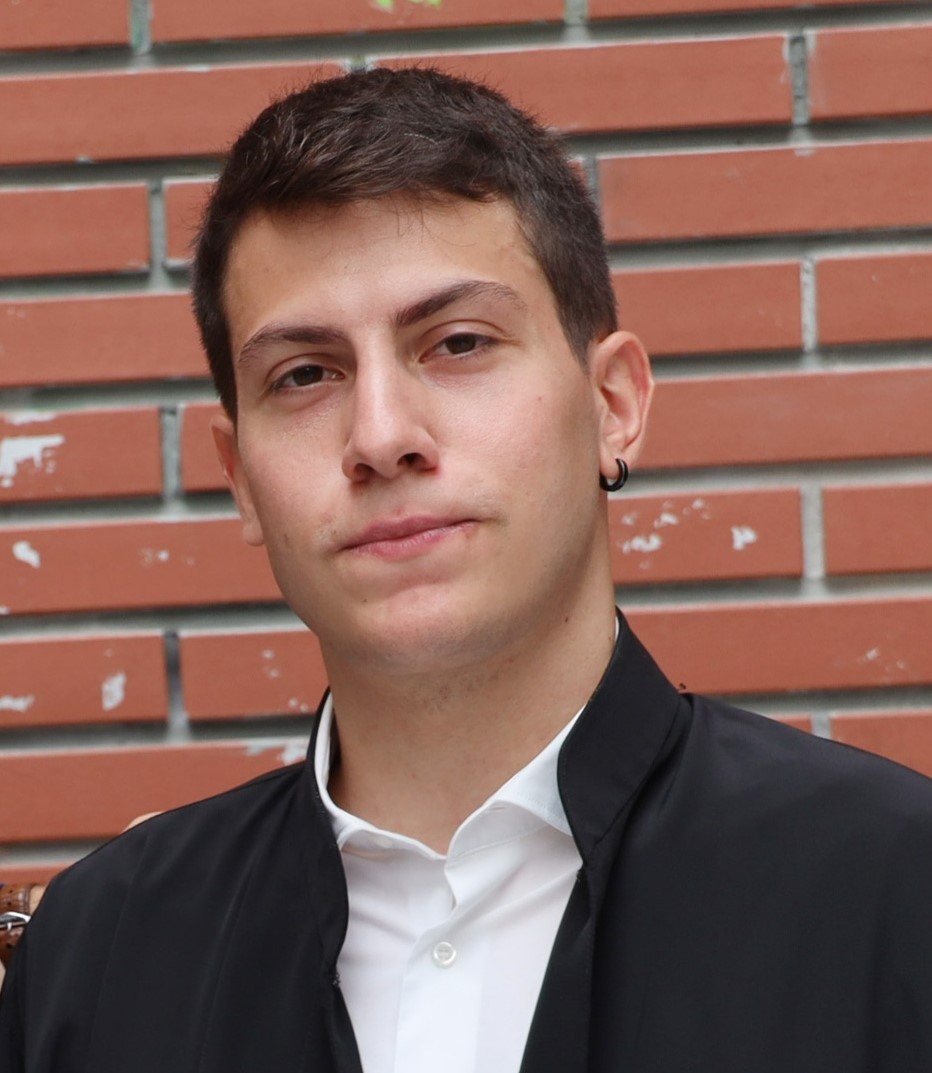}}]{Theocharis Apostolakis} received the Dipl.Ing. degree in Mechanical Engineering from the University of Thessaly, Greece, in 2022. He is currently a PhD student with the same department. His research interests include Control Systems and Optimization.   
\end{IEEEbiography}

\begin{IEEEbiography}[{\includegraphics[width=1in,height=1.25in,clip,keepaspectratio]{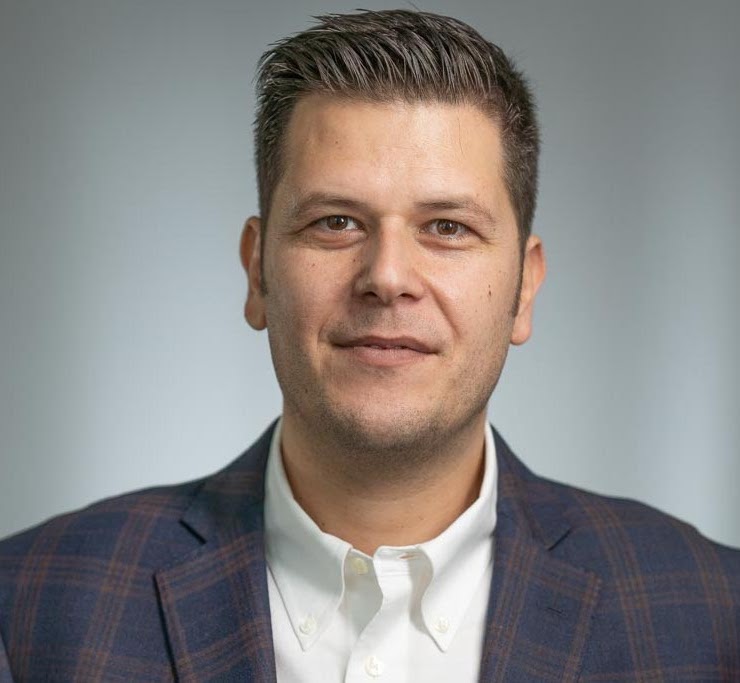}}]{Michail Makridis} 
is a Senior Research Scientist for the Institute of Transport Planning of the Department of Civil, Environmental and Geomatic Engineering, ETH Z{\" u}rich. He holds a Ph.D. in Computer Vision from the Democritus University of Thrace, Greece. From 2012 until 2015, he was working for the European Commission Joint Research Centre as a Postdoc researcher on Data mining for Maritime Transport datasets. From 2016 until 2020 he was the scientific responsible for the Traffic Modeling Group of the Sustainable Transport Unit of the Directorate of Energy, Transport and Climate, supporting European Policies on Intelligent Transportation Systems (ITS) and highly Automated Vehicles (AVs). His interests are in ITS, Traffic Management and Simulation, AVs, Vehicle Dynamics, and Driver Behavior, Energy demand and Emissions, and Computer Vision. He is a member of IEEE.
\end{IEEEbiography}

\begin{IEEEbiography}[{\includegraphics[width=1in,height=1.25in,clip,keepaspectratio]{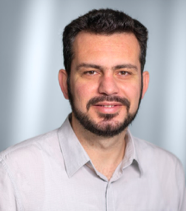}}]{Anastasios Kouvelas} 
Anastasios Kouvelas is the Director of the Traffic Engineering and Control research group at the Institute for Transport Planning and Systems (IVT), Dept. of Civil, Environmental and Geomatic Engineering, ETH Zurich (since August 2018). Prior to joining IVT, he was a Research Scientist at the Urban Transport Systems Laboratory (LUTS), EPFL (2014-2018), and a Postdoctoral Fellow with Partners for Advanced Transportation Technology (PATH) at the University of California, Berkeley (2012–2014).
\end{IEEEbiography}

\begin{IEEEbiography}[{\includegraphics[width=1in,height=1.25in,clip,keepaspectratio]{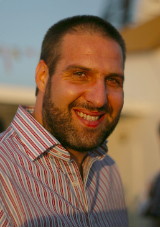}}]{Konstantinos Ampountolas}
(Member, IEEE) received the Dipl.Ing. degree in production engineering and management, the M.Sc. degree in operations research, and the Ph.D. degree in engineering from the Technical University of Crete, Greece, in 1999, 2002, and 2009, respectively. He was a Senior Lecturer with the James Watt School of Engineering, University of Glasgow, U.K., from 2013 to 2019, a Research Fellow with the {\' E}cole Polytechnique F{\' e}d{\' e}rale de Lausanne, Switzerland, from 2012 to 2013, a Visiting Researcher Scholar with the University of California at Berkeley, Berkeley, CA, USA, in 2011, and a Post-Doctoral Researcher with the Centre for Research \& Technology Hellas, Greece, in 2010. He was also a short-term Visiting Professor with the Technion–Israel Institute of Technology, Israel, in 2014, and the Federal University of Santa Catarina, Florianópolis, Brazil, in 2016 and 2019. Since 2019, he has been an Associate Professor with the Department of Mechanical Engineering, University of Thessaly, Greece. His research interests include control and optimization with applications to transport networks and systems.

He has served as the Editor for \emph{Transportation} of \emph{Data in Brief}, from 2018 to 2019, as an Associate Editor for the \emph{Journal of Big Data Analytics in Transportation}, from 2018 to 2020, and on the editorial advisory boards of \emph{Transportation Research Part C} (from 2014 to 2021) and \emph{Transportation Research Procedia} (since 2014).
\end{IEEEbiography}
\vfill
\end{document}